\documentclass[aps,prd,onecolumn,notitlepage,nofootinbib,showpacs,preprintnumbers,superscriptaddress,groupedaddress]{revtex4-1}
\usepackage[T1]{fontenc}
\usepackage[utf8]{inputenc}
\usepackage{bm}
\usepackage{amsmath}
\usepackage{amssymb}
\usepackage{multirow}
\usepackage{esint}
\usepackage{xcolor}
\usepackage{array}
\usepackage{float}
\usepackage{graphicx}
\PassOptionsToPackage{normalem}{ulem}
\usepackage{ulem}
\usepackage{hyperref}

\begin{document}
	
	\title{Higher derivative scalar-tensor theory from the spatially covariant gravity:\\a linear algebraic analysis}
	
	\author{Xian Gao}%
	\email[Email: ]{gaoxian@mail.sysu.edu.cn}
	\affiliation{%
		School of Physics and Astronomy, Sun Yat-sen University, Guangzhou 510275, China}
	
	\date{June 28, 2020}
	
	\begin{abstract}
		We investigate the ghostfree scalar-tensor theory with a timelike scalar field, with derivatives of the scalar field up to the third order and with the Riemann tensor up to the quadratic order. We build two types of linear spaces. One is the set of linearly independent generally covariant scalar-tensor monomials, the other is the set of linearly independent spatially covariant gravity monomials. We argue that these two types of linear space are isomorphic to each other in the sense of gauge fixing/recovering procedures. We then identify the subspaces in the spatially covariant gravity, which are spanned by linearly independent monomials built of the extrinsic and intrinsic curvature, the lapse function as well as their spatial derivatives, up to the fourth order in the total number of derivatives.	The vectors in these subspaces, i.e., spatially covariant polynomials, automatically propagate at most three degrees of freedom. As a result, their images under the gauge recovering mappings are automatically the subspaces of scalar-tensor theory that propagate up to three degrees of freedom as long as the scalar field is timelike. The mappings from the spaces of spatially covariant gravity to the spaces of scalar-tensor theory are encoded in the projection matrices, of which we also derived the expressions explicitly. Our formalism and results can be useful in deriving the generally covariant higher derivative scalar-tensor theory without ghost(s).
	\end{abstract}
	
	\maketitle

\section{Introduction}

Killing the ghost(s), i.e., evading the ghostlike instabilities or simply the unwanted modes, is one of the central problems in the developments of modified gravity in the past decades.
The scalar-tensor theory is introduced as one of the main theories of modified gravity, which introduces additional scalar degree(s) of freedom (DoF) other than the two tensorial DoFs of the General Relativity.
When higher order derivatives of the scalar field or higher curvature terms are present, the instabilities called the Ostrogradsky ghost(s) generally arise \cite{Woodard:2015zca}. 
Thus one of the question is how to introduce higher derivatives and/or higher curvature terms without ghost(s). 
The representative achievements are known to be the Horndeski theory \cite{Horndeski:1974wa,Nicolis:2008in,Deffayet:2011gz,Kobayashi:2011nu} as well as the degenerate higher-order derivative scalar tensor theory \cite{Gleyzes:2014dya,Gleyzes:2014qga,Langlois:2015cwa,Motohashi:2016ftl} (see \cite{Langlois:2018dxi,Kobayashi:2019hrl} for reviews), which include derivatives of the scalar field up to the second order and the Riemann curvature tensor up to the linear order.

On the other hand, it is necessary and possible to go beyond by introducing derivatives of the scalar field higher than the second order and curvature terms beyond the linear order \cite{Gao:2020juc,Gao:2020yzr}.
The key observation is that the third and higher order derivatives of the scalar field are of the same importance as the higher curvature terms and thus should be consider together.
Moreover, it is possible that novel ghostfree Lagrangians will arise by combining the higher derivatives of the scalar field and higher curvature terms (with couplings of the scalar field).
Another motivation comes from the phenomenological side. As the parameter spaces of the Horndeski-like theories are highly restricted, e.g., after taking into account the constraint of the propagation speed of the gravitational waves \cite{Creminelli:2017sry,Sakstein:2017xjx,Ezquiaga:2017ekz,Baker:2017hug,Amendola:2017orw,Langlois:2017dyl} (see \cite{Ezquiaga:2018btd} for a review), one may wonder if scalar-tensor theories with even higher order derivatives and higher curvature terms can supply us a broader playground that may pass the observational tests \cite{Gao:2019liu}.

Indeed, a generic and straightforward approach to go to higher orders is to finely tune the structure of the higher derivatives as well as the couplings between curvature and the scalar field such that they are degenerate, which is similar to the construction of degenerate higher derivative scalar-tensor theories. 
Although it has been explored in the case of point particles \cite{Motohashi:2017eya,Motohashi:2018pxg}, the generalization to the field theory is still missing. 
There is an alternative approach to the ghostfree scalar-tensor theory beyond the second order in derivatives and the linear order in the curvature tensor, which is inspired by the Chern-Simons gravity \cite{Lue:1998mq,Jackiw:2003pm}, the ghostfree Weyl gravity \cite{Deruelle:2012xv} as well as the recently studied ghostfree quadratic gravity with parity violation \cite{Crisostomi:2017ugk}. These theories generally propagates ghostlike DoFs, but these ghost modes can be eliminated (or are invisible) when the gradient of the scalar field is timelike\footnote{In this work we sometimes refer to this property as ``the scalar field is timelike'' or ``a timelike scalar field'' for short.}. 
Although the configuration of the scalar field has to be restricted to be timelike, i.e., these theories can be argued to be ghostfree only in part of their full phase space, these theories indicate the possible existence of ghostfree scalar-tensor theories with derivatives beyond the second order and with curvature tensor beyond the linear order.

When being written in coordinates that are adapted with the foliation structure\footnote{These are exactly the Arnowitt-Deser-Misner (ADM) coordinates. Sometimes choosing the ADM coordinates is also referred to fixing the so-called ``unitary gauge'', in which the scalar field is chosen to be spatially uniform.}, generally covariant scalar-tensor theory (GST) with a timelike scalar field can be recast to be a pure metric theory with only spatial diffeomorphism, which we may dub as the spatially covariant gravity (SCG).
The well-studied effective field theory (EFT) of inflation \cite{Creminelli:2006xe,Cheung:2007st} and its generalization to the EFTs of dark energy \cite{Gubitosi:2012hu,Gleyzes:2013ooa} (see e.g., \cite{Tsujikawa:2014mba,Frusciante:2019xia} for reviews), of non-singular cosmology \cite{Cai:2016thi,Cai:2017tku}, of $f(T)$ gravity \cite{Li:2018ixg},  as well as the Ho\v{r}ava gravity \cite{Horava:2009uw,Blas:2009qj} (including the scalar Einstein-Aether theory \cite{Jacobson:2014mda}) are examples of SCG theories.
The GST and SCG can be viewed as the two faces of the same theory, as long as the the scalar field is timelike.
Although it might be involved to construct ghostfree higher derivative GST directly, in the framework of SCG, however, it is relatively straightforward to build the theory with at most three degrees of freedom  \cite{Gao:2014soa,Gao:2014fra,Fujita:2015ymn,Gao:2018znj,Gao:2019lpz,Gao:2018izs,Gao:2019twq}.
We may thus use the SCG as the ``generator'' of ghostfree higher derivative GST, in particular, with derivatives of the scalar field beyond the second order and curvature tensor beyond the linear order.

There is a one-to-one correspondence between a GST term and a SCG term, through the gauge-fixing and gauge-recovering (Stueckelberg trick) procedures. 
In this work we concentrate on the terms that are polynomials. 
Precisely, for the GST terms, we consider polynomials built of the scalar field as well as the Riemann curvature tensor together with their generally covariant derivatives. 
For the SCG terms, we consider polynomials built of the extrinsic and intrinsic curvature, the lapse function together with their intrinsic (spatial) and extrinsic (Lie) derivatives.
Previously, we have classified the GST monomials in \cite{Gao:2020juc}. 
For the SCG, we have exhausted and classified the monomials in \cite{Gao:2020yzr} with only spatial derivatives.
SCG with only spatial derivatives is a well-defined framework, which automatically evades the unwanted ghostlike mode.
In other words, linearly independent monomials of SCG with only spatial derivatives define a ``linear space'', in which SCG polynomials with only spatial derivatives can be viewed as ``vectors''.
On the other hand, linearly independent GST monomials also form a linear space.
Since the two types of monomials have one-to-one correspondence, we may simply map the linear space of SCG monomials to the linear space of GST monomials, such that the latter is a subspace of the full space of GST monomials, in which the ``vectors'' are GST polynomials that are ghostfree (in the sense that there are at most 3 DoFs) as long as the scalar field is timelike.
In particular, such ``image'' subspaces of GST can be used as the starting point, which is more convenient than the full space of GST, to construct the ghostfree covariant higher derivative scalar-tensor theories.

This work is devoted to a systematic linear algebraic analysis of the correspondence between SCG and GST.
The rest of the paper is organized as following.
In Sec. \ref{sec:ls}, we classify the monomials of both the generally covariant scalar-tensor theory and the spatially covariant gravity, and pay special attention to their linear algebraic structures.
In Sec. \ref{sec:d123} and Sec. \ref{sec:d4}, we study the maps from the space of spatially covariant gravity monomials to the space of generally covariant scalar-tensor monomials.
Sec. \ref{sec:con} concludes.

\section{A linear space consideration} \label{sec:ls}

In this section, we make a detailed classification of two types of monomials. 
\begin{itemize}
	\item One is the set of generally covariant scalar-tensor (GST) monomials, which are built of the generally covariant derivatives of the scalar field up to the third order and their coupling with the Riemann curvature tensor up to the quadratic order.
	\item The other is the set of spatially covariant gravity (SCG) monomials, which are built of the extrinsic and intrinsic curvature, the lapse function as well as their spatial derivatives, with total number of derivative no larger than four.
\end{itemize}
The key point of view we shall take is that these two sets of monomials are linear spaces, and the gauge fixing/recovering process are linear maps.

\subsection{A linear space consideration}

\subsubsection{Space of generally covariant scalar-tensor monomials}

A general GST monomial take the form
	\begin{equation}
	\underbrace{\cdots {}^{4}\! R\cdots}_{c_{0}}\underbrace{\cdots\nabla \,{}^{4}\! R\cdots}_{c_{1}}\underbrace{\cdots\nabla\nabla \,{}^{4}\! R\cdots}_{c_{2}}\cdots\underbrace{\cdots\nabla\nabla\phi\cdots}_{d_{2}}\underbrace{\cdots\nabla\nabla\nabla\phi\cdots}_{d_{3}}\underbrace{\cdots\nabla\nabla\nabla\nabla\phi\cdots}_{d_{4}}\cdots, \label{STmonoform}
	\end{equation}
where ``$\cdots$'' denotes multiple Riemann curvature tensor (we schematically denote as ${}^{4}\!R$), the scalar field and their covariant derivatives $\nabla_{a}$.
All the 4-dimension indices are contracted by the spacetime metric $g^{ab}$, the first derivative of the scalar field $\nabla^{a}\phi$ and/or the completely antisymmetric Levi-Civita tensor $\varepsilon^{abcd}$. 
One can imagine that there are various types of GST monomials and their total number is huge.
In order to classify the monomials, we assign each GST monomial a set of integers $\left(c_{0},c_{1},c_{2},\cdots;d_{2},d_{3},d_{4},\cdots\right)$, where $c_{0},c_{1},c_{2},\cdots$ are the numbers of Riemann curvature tensor and its first, second derivatives, etc., $d_{2},d_{3},d_{4},\cdots$ are the numbers the second, the third and the fourth covariant derivatives of $\phi$, etc..
We assume all the $c_{n}$'s and $d_{n}$'s are non-negative integers.	
It is convenient to define a single number \cite{Gao:2020juc}
	\begin{equation}
	d \equiv \sum_{n=0}\left[\left(n+2\right)c_{n}+\left(n+1\right)\,d_{n+2}\right], \label{d_ST_def}
	\end{equation}
to characterize the overall ``order'' of derivatives of each monomial.
In fact, $d$ is nothing but the total number of derivatives of the SCG terms corresponding to the given GST monomial. 
Using the integer $d$ makes the correspondence between the scalar-tensor monomials and the spatially covariant gravity monomials transparent.

As being argued in Ref. \cite{Gao:2020juc}, possible ghostfree GST terms can arise only from the combinations of GST monomial of the same order $d$, i.e., GST polynomials of order $d$.
Thus we shall group various GST monomials according to the integer $d$, and study them order by order.
For each $d$, one is able to build a ``complete basis'' for the GST monomials, which is a set of linearly independent GST monomials, such that any GST polynomial of order $d$ is a \emph{linear combination} of the monomials in the complete basis, of which the coefficients are functions of $\phi$ and $\nabla\phi$. 
The monomials in the complete basis are chosen to be algebraically linear independent\footnote{In this work, we consider only algebraic linear (in)dependence. Clearly there are also linear dependences through total derivatives, which should be taken into account when building the Lagrangians.}.
In this sense, the complete basis of order $d$ spans a linear space, which we dub as $\Sigma_{d}$. Any GST polynomial of order $d$ can thus be viewed as a vector in this linear space.

With this picture in mind, we may further introduce two subspace of $\Sigma_{d}$ at each order $d$.
One is the ghost-free combination of the monomials that we dub as $\Omega_{d}$, the other is the set of monomials that are ghost-free as long as the scalar field is timelike, which we dub as $\Gamma^{\ast}_{d}$.
It is clear that
	\begin{equation}
		\Sigma_{d} \supset \Gamma^{\ast}_{d} \supset \Omega_{d}.
	\end{equation}

\subsubsection{Space of spatially covariant gravity monomials}

A similar investigation of SCG monomials was performed \cite{Gao:2020yzr}.
A general SCG monomial is built of the spatial (intrinsic) curvature ${}^{3}\!R_{ij}$, the extrinsic curvature $K_{ij}$, the lapse function $N$, together with their spatial derivatives $\mathrm{D}_{i}$ as well as the Lie derivative $\pounds_{\bm n}$.
The spatial derivative $\mathrm{D}_{i}$ is compatible with the spatial metric, which is thus the intrinsic derivative on the spatial hypersurfaces. 
The Lie derivative $\pounds_{\bm n}$ is defined to be with respect to the normal vector $\bm{n}$ of the spatial hypersurfaces, which acts as the extrinsic or temporal derivative in the framework of SCG. Similar to (\ref{STmonoform}), a general SCG monomial takes the form
	\begin{equation}
		\underbrace{\cdots\,{}^{3}\!R\cdots}_{r_{0}}\underbrace{\cdots\Delta\,{}^{3}\!R\cdots}_{r_{1}}\cdots\underbrace{\cdots K\cdots}_{k_{0}}\underbrace{\cdots\Delta K\cdots}_{k_{1}}\cdots\underbrace{\cdots\Delta N\cdots}_{l_{1}}\underbrace{\cdots\Delta^{2}N\cdots}_{l_{2}}\cdots, \label{SCGmonoform}
	\end{equation}
where $\,{}^{3}\!R$, $K$ schematically denote the spatial and the extrinsic curvature, respectively.
In (\ref{SCGmonoform}), $\Delta$ stands for both the spatial derivative and the Lie derivative as well as their mixings, since generally both types of derivatives are allowed.
The indices in (\ref{SCGmonoform}) are purely spatial, which runs from $1$ to $3$, are summed by the spatial metric $h^{ij}$ as well as the spatial Levi-Civita tensor $\varepsilon^{ijk}$.
The integers $r_{m}$, $k_{n}$ and $l_{p}$, which are assumed to be non-negative, are the numbers of the $m$-th, the $n$-th and the $p$-th spatial/Lie derivatives of ${}^{3}\!R_{ij}$, $K_{ij}$, and $N$, respectively.

We can also assign a single integer to characterize the overall order of derivatives of each SCG monomial.
It is clear that in the framework of SCG, $K_{ij}$ is of $\mathcal{O}(\Delta^{1})$ and $\,{}^{3}\!R_{ij}$ is of $\mathcal{O}(\Delta^{2})$.
As a result, the overall order of derivatives of the monomial is
	\begin{equation}
		d=\sum_{n=0}\left[\left(n+2\right)r_{n}+\left(n+1\right)\left(k_{n}+l_{n+1}\right)\right] . \label{d_SCG_def}
	\end{equation}
It is interesting that (\ref{d_SCG_def}) takes exactly the same form as (\ref{d_ST_def}), which also shows the advantage of using $d$ as the label of orders of monomials in both GST and SCG.
We have deliberately defined $d$ in (\ref{d_ST_def}) such that the same number of $d$ can be used to characterize the overall order of derivative for both the GST and SCG monomials.
For this reason, we use the same symbol $d$ in (\ref{d_ST_def}) and (\ref{d_SCG_def}).

Similar to the linear space $\Sigma_{d}$, the set of linearly independent SCG monomials of the same order $d$ spans a linear space, which we dub as $\mathcal{S}_{d}$.
A general polynomial, which is the linear combination of SCG monomials of order $d$ with coefficients being functions of $t$ and $N$, can thus be viewed as a vector in $\mathcal{S}_{d}$.

Generally, when the Lie derivatives are present, there are unwanted extra modes propagating.
On the other hand, it has been proved that the total number of DoFs is no large than 3 if only the spatial derivatives are present.
Thus we may restrict ourselves in the subspace of $\mathcal{S}_{d}$ with only the spatial derivatives are present. We refer to this subspace as $\mathcal{G}_{d}$.
We should emphasize that $\mathcal{G}_{d}$ is by no means the largest subspace of $\mathcal{S}_{d}$ that propagate at most 3 DoFs, which we may dub as $\mathcal{G}^{\ast}_{d}$.
In fact, in the case with $\pounds_{\bm{n}}N$, it is possible to tune the coefficients of monomials such that the combination (i.e., SCG polynomial) is degenerate and there is no unwanted extra mode \cite{Gao:2018znj}.
Nevertheless, the subspace $\mathcal{G}_{d}$ is well-defined and thus we shall focus on this subspace\footnote{It has been shown that for some specific examples, terms with $\pounds_{\bm{n}}N$ that propagate no extra mode can be related to terms without $\pounds_{\bm{n}}N$ by field transformations \cite{Gao:2019lpz}. This indicates the possibility that the whole subspace $\mathcal{G}^{\ast}_{d}$ may be mapped to $\mathcal{G}_{d}$ under field transformations, although a full investigation is still required.}.

By using the Stueckelberg trick, the SCG terms can be mapped to the GST terms. The explicit expressions have been given in Ref. \cite{Gao:2020yzr}. Accordingly, we can assign each SCG term the set of integers $(c_{0},c_{1},\cdots;d_{2},d_{3},\cdots)$ of the corresponding GST terms.
In particular, we shall use the following correspondences:
	\begin{eqnarray}
	K_{ij}\sim a_{i} & \sim & \left(0;1,0\right), \label{coord_1}\\
	{}^{3}\!R_{ij} & \sim & \left(1;0,0\right) \sim \left(0;2,0\right) ,\\
	\mathrm{D}_{k}K_{ij}\sim\mathrm{D}_{j}a_{i} & \sim & \left(0;0,1\right), \label{coord_3}
	\end{eqnarray}
where only the set of 3 integers $(c_{0};d_{2},d_{3})$ are present.

\subsubsection{Correspondences}

By making use the gauge fixing/recovering procedures, the GST and SCG terms can be transformed to each other. 
Precisely, a GST monomial of order $d$ will be transformed to a SCG polynomial which is a combination of SCG monomials of order $d$, and vice versa. 
In the language of linear spaces, there is a one-to-one correspondence between a vector in the GST space  $\Sigma_{d}$ and a vector in the SCG space $\mathcal{S}_{d}$. 
In other words, the linear spaces of the GST monomials and SCG monomials of the same order $d$ are isomorphic to each other, which also implies
\begin{equation}
\dim (\Sigma_{d}) = \dim (\mathcal{S}_{d}).
\end{equation}
Note this provides a more precise formulation of the statement that the GST and SCG are two faces of the same theory, in the sense that they are two different representations (set of basis) of the same linear space.

\begin{figure}[H]
	\begin{center}
		\includegraphics[scale=0.4]{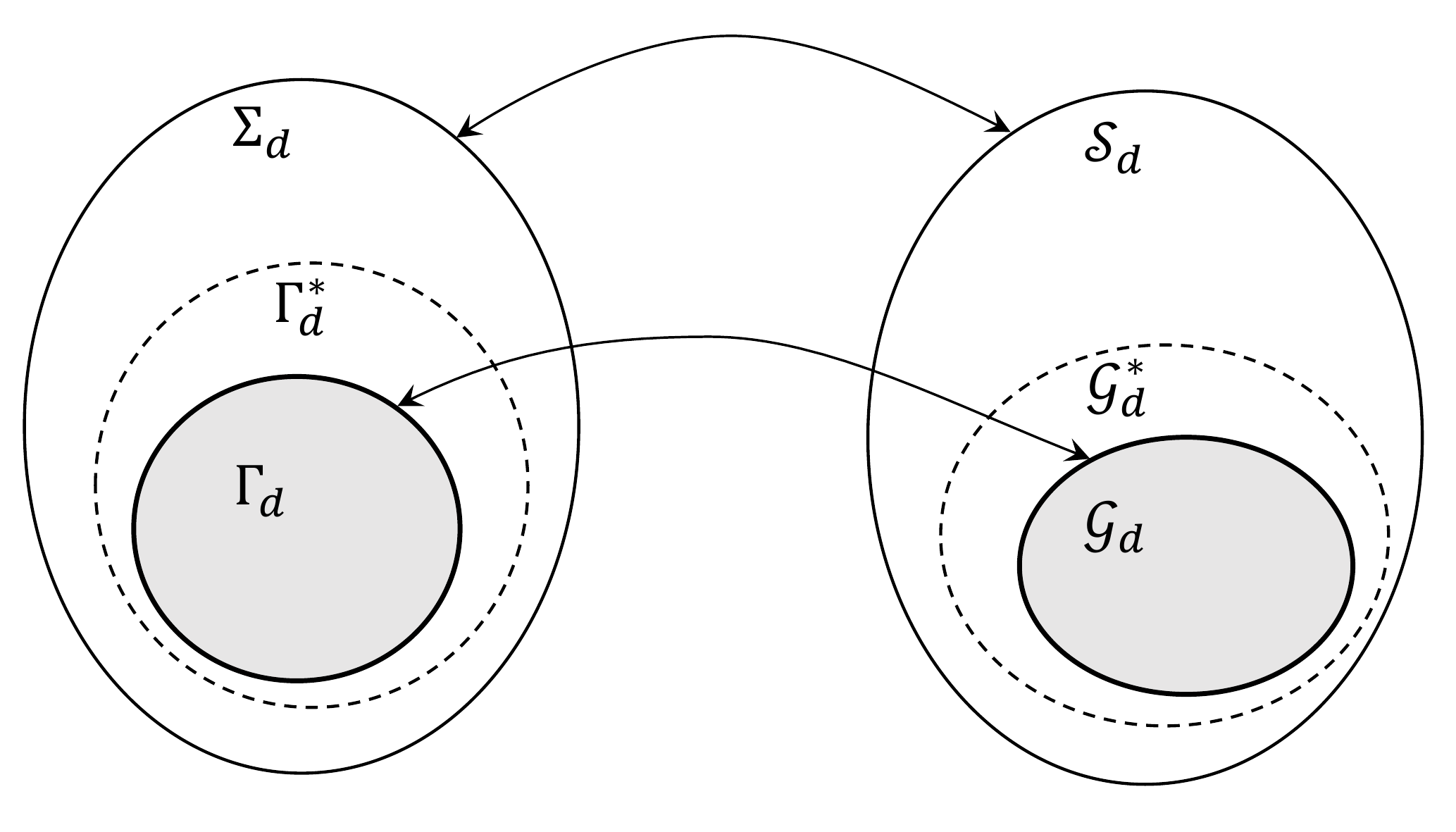}
		\par\end{center}
	\caption{Spaces and maps of the GST and SCG monomials.}
	\label{fig:maps}
\end{figure}

Correspondingly, the subspace $\mathcal{G}_{d}$ is mapped to a subspace of $\Sigma_{d}$, which we dub as $\Gamma_{d}$.
Since the subspace $\mathcal{G}_{d}$ propagates at most 3 DoFs, its image automatically picks out the subspace $\Gamma_{d}$ of GST that propagate at most 3 DoFs as long as the scalar field is timelike.
The main task of this work, is thus to identify the subspace $\Gamma_{d}$.
Clearly $\Gamma_{d}$ is much ``smaller'' than the full space $\Sigma_{d}$, which is thus can be used as a starting point of searching the final ghostfree GST.
The various subspaces and maps are illustrated in Fig. \ref{fig:maps}. 

At each order $d$, we denote the basis of the space of SCG monomials $\mathcal{S}_{d}$ as $\left\{ \bm{s}_{a}^{(d)}\right\}$ with $a=1,\cdots,\dim(\mathcal{S}_{d})$, and the basis of the subspace $\mathcal{G}_{d}$ as $\left\{ \bm{g}_{a}^{(d)}\right\}$ with $a=1,\cdots,\dim(\mathcal{G}_{d})$.
For the space of GST monomials, we the basis of $\Sigma_{d}$ as $\left\{ \bm{\sigma}_{\alpha}^{(d)}\right\} $ with $\alpha = 1,\cdots,\dim(\Sigma_{d})$.
In this work, we shall choose $\bm{g}_{a}^{(d)}$ as SCG monomials and $\bm{\sigma}_{\alpha}^{(d)}$ as GST monomials, respectively.
In some sense we may refer to such basis as the ``natural'' or simple basis.
From the point of view of linear spaces, one is free to choose other independent basis, such as polynomials.
After the gauge recovering mapping (Stueckelberg trick), we have
	\begin{equation}
	\left\{ \bm{g}_{a}^{(d)}\right\} \rightarrow \left\{ \tilde{\bm{g}}_{a}^{(d)}\right\} , \quad a=1,\cdots,\dim(\mathcal{G}_{d}),
	\end{equation}
where $\tilde{\bm{g}}_{a}^{(d)}$ is the image of $\bm{g}_{a}^{(d)}$ for each $a$. Note each $\bm{g}_{a}^{(d)}$, which is a SCG monomial, is mapped to a GST term that is generally a linear combination of GST monomials of order $d$, i.e., a GST polynomial. 
On the other hand, since the complete basis of GST polynomials of order $d$ is $\left\{ \bm{\sigma}_{\alpha}^{(d)}\right\} $, we can thus write
	\begin{equation}
	\tilde{\bm{g}}_{a}^{(d)}=\sum_{\alpha = 1}^{\dim (\Sigma_{d})}\Phi_{a\alpha}^{(d)}\bm{\sigma}_{\alpha}^{(d)}, \qquad a= 1,\cdots,\dim(\mathcal{G}_{d}) . \label{Phimat_def}
	\end{equation}
It is illuminating to dub  $\Phi_{a\alpha}^{(d)}$ as the ``projection matrix'', as it projects $\Sigma_{d}$ to its subspace $\Gamma_{d}$.
The set of $\left\{ \tilde{\bm{g}}_{a}^{(d)}\right\}$ is nothing but spans the desired subspace $\Gamma_{d}$, i.e., can be viewed as the basis of $\Gamma_{d}$.

For the parity-violating case, we can draw a figure similar to Fig. \ref{fig:maps}. 
The discussion is also parallel to that of the parity-preserving case. For the spaces of SCG monomials, we consider the subspace of linear independent parity-violating SCG monomials with only spatial derivatives, which is the counterpart of $\mathcal{G}_{d}$ and we denote as $\mathcal{Z}_{d}$. The basis of $\mathcal{Z}_{d}$ are denoted as $\left\{ \bm{z}_{a}^{(d)}\right\} $ with $a= 1,\cdots,\dim(\mathcal{Z}_{d})$.
For the space of GST monomials, we denote the full space of parity-violating GST monomials to be $\Xi_{d}$, of which the basis are $\left\{ \bm{\xi}_{\alpha}^{(d)}\right\} $ with $\alpha = 1,\cdots,\dim(\Xi_{d})$.
After making the gauge recovering mapping, we have
	\begin{equation}
	\left\{ \bm{z}_{a}^{(d)}\right\} \rightarrow \left\{ \tilde{\bm{z}}_{a}^{(d)}\right\} , \quad a= 1,\cdots,\dim(\mathcal{Z}_{d}),
	\end{equation}
where $\tilde{\bm{z}}_{a}^{(d)}$ are parity-violating GST polynomials for each $a$.
Similar to (\ref{Phimat_def}), $\tilde{\bm{z}}_{a}^{(d)}$ can be expanded as
	\begin{equation}
	\tilde{\bm{z}}_{a}^{(d)}=\sum_{\alpha = 1}^{\dim(\Xi_{d})} \Psi_{a\alpha}^{(d)} \bm{\xi}_{\alpha}^{(d)}, \qquad a= 1,\cdots,\dim(\mathcal{Z}_{d}) , \label{Psimat_def}
	\end{equation}
where we also define the projection matrix $\Psi_{a\alpha}^{(d)}$ for the parity-violating case.
The set of $\left\{ \tilde{\bm{z}}_{a}^{(d)}\right\}$  spans the subspace ${\Theta}_{d}$, which is the counterpart of $\Gamma_{d}$.

From the above discussion, the subspaces $\Gamma_{d}$ and ${\Theta}_{d}$,  which are spaces of GST polynomials that are ghost-free as long as the scalar field is timelike, are completely encoded in the projection matrices $\Phi_{a\alpha}^{(d)}$ and $\Psi_{a\alpha}^{(d)}$.
The main task in the rest of this work is thus to derive the GST polynomials $\tilde{\bm{g}}_{a}^{(d)}$ and $\tilde{\bm{z}}_{a}^{(d)}$ by using the Stueckelberg trick, and show the explicit expressions for the projection matrices $\Phi_{a\alpha}^{(d)}$ and $\Psi_{a\alpha}^{(d)}$.

\subsection{GST monomials} \label{sec:STmono}

The classification of the GST monomials and the complete basis for each $d$ up to $d=4$ have been developed in \cite{Gao:2020juc}. 
Here we reformulate the results for the purpose of the present work, with improved notations.

\subsubsection{Parity preserving}

We exhaust all the GST monomials up to $d=4$.
The results are summarized in Tab. \ref{tab:STbasis}. 
	\begin{center}
		\begin{table}[H]
			\begin{centering}
				\begin{tabular}{|c|c|l|>{\raggedright}p{7cm}|}
					\hline 
					$d$ & $\left(c_{0};d_{2},d_{3}\right)$ & Unfactorizable & Factorizable\tabularnewline
					\hline 
					\hline 
					$0$ & $\left(0;0,0\right)$ & $\mathcal{E}^{\left(0;0,0\right)}$ & -\tabularnewline
					\hline 
					\hline 
					$1$ & $\left(0;1,0\right)$ & $\mathcal{E}^{\left(0;1,0\right)}$ & -\tabularnewline
					\hline 
					\hline 
					\multirow{2}{*}{$2$} & $\left(0;2,0\right)$ & $\mathcal{E}^{\left(0;2,0\right)}$ & $\mathcal{E}^{\left(0;1,0\right)}\otimes\mathcal{E}^{\left(0;1,0\right)}$\tabularnewline
					\cline{2-4} \cline{3-4} \cline{4-4} 
					& $\left(1;0,0\right)$ & $\mathcal{E}^{\left(1;0,0\right)}$ & -\tabularnewline
					\hline 
					\hline 
					\multirow{3}{*}{$3$} & $\left(0;3,0\right)$ & $\mathcal{E}^{\left(0;3,0\right)}$ & $\mathcal{E}^{\left(0;1,0\right)}\otimes\mathcal{E}^{\left(0;2,0\right)}$
					
					$\mathcal{E}^{\left(0;1,0\right)}\otimes\mathcal{E}^{\left(0;1,0\right)}\otimes\mathcal{E}^{\left(0;1,0\right)}$\tabularnewline
					\cline{2-4} \cline{3-4} \cline{4-4} 
					& $\left(0;1,1\right)$ & $\mathcal{E}^{\left(0;1,1\right)}$ & $\mathcal{E}^{\left(0;1,0\right)}\otimes\mathcal{E}^{\left(0;0,1\right)}$\tabularnewline
					\cline{2-4} \cline{3-4} \cline{4-4} 
					& $\left(1;1,0\right)$ & $\mathcal{E}^{\left(1;1,0\right)}$ & $\mathcal{E}^{\left(1;0,0\right)}\otimes\mathcal{E}^{\left(0;1,0\right)}$\tabularnewline
					\hline 
					\hline 
					\multirow{6}{*}{$4$} & $\left(0;4,0\right)$ & $\mathcal{E}^{\left(0;4,0\right)}$ & $\mathcal{E}^{\left(0;1,0\right)}\otimes\mathcal{E}^{\left(0;3,0\right)}$
					
					$\mathcal{E}^{\left(0;2,0\right)}\otimes\mathcal{E}^{\left(0;2,0\right)}$
					
					$\mathcal{E}^{\left(0;1,0\right)}\otimes\mathcal{E}^{\left(0;1,0\right)}\otimes\mathcal{E}^{\left(0;2,0\right)}$
					
					$\mathcal{E}^{\left(0;1,0\right)}\otimes\mathcal{E}^{\left(0;1,0\right)}\otimes\mathcal{E}^{\left(0;1,0\right)}\otimes\mathcal{E}^{\left(0;1,0\right)}$\tabularnewline
					\cline{2-4} \cline{3-4} \cline{4-4} 
					& $\left(0;2,1\right)$ & $\mathcal{E}^{\left(0;2,1\right)}$ & $\mathcal{E}^{\left(0;1,0\right)}\otimes\mathcal{E}^{\left(0;1,1\right)}$
					
					$\mathcal{E}^{\left(0;2,0\right)}\otimes\mathcal{E}^{\left(0;0,1\right)}$
					
					$\mathcal{E}^{\left(0;1,0\right)}\otimes\mathcal{E}^{\left(0;1,0\right)}\otimes\mathcal{E}^{\left(0;0,1\right)}$\tabularnewline
					\cline{2-4} \cline{3-4} \cline{4-4} 
					& $\left(0;0,2\right)$ & $\mathcal{E}^{\left(0;0,2\right)}$ & $\mathcal{E}^{\left(0;0,1\right)}\otimes\mathcal{E}^{\left(0;0,1\right)}$\tabularnewline
					\cline{2-4} \cline{3-4} \cline{4-4} 
					& $\left(1;2,0\right)$ & $\mathcal{E}^{\left(1;2,0\right)}$ & $\mathcal{E}^{\left(1;1,0\right)}\otimes\mathcal{E}^{\left(0;1,0\right)}$
					
					$\mathcal{E}^{\left(1;0,0\right)}\otimes\mathcal{E}^{\left(0;2,0\right)}$
					
					$\mathcal{E}^{\left(1;0,0\right)}\otimes\mathcal{E}^{\left(0;1,0\right)}\otimes\mathcal{E}^{\left(0;1,0\right)}$\tabularnewline
					\cline{2-4} \cline{3-4} \cline{4-4} 
					& $\left(2;0,0\right)$ & $\mathcal{E}^{\left(2;0,0\right)}$ & $\mathcal{E}^{\left(1;0,0\right)}\otimes\mathcal{E}^{\left(1;0,0\right)}$\tabularnewline
					\cline{2-4} \cline{3-4} \cline{4-4} 
					& $\left(1;0,1\right)$ & $\mathcal{E}^{\left(1;0,1\right)}$ & $\mathcal{E}^{\left(1;0,0\right)}\otimes\mathcal{E}^{\left(0;0,1\right)}$\tabularnewline
					\hline 
				\end{tabular}
				\par\end{centering}
			\caption{Classification of the parity-preserving GST monomials.}
			\label{tab:STbasis}
		\end{table}
		\par\end{center}
We comment on the construction of the complete basis by explaining how we make Tab. \ref{tab:STbasis}.
	\begin{itemize}
		\item First, we classify the GST monomials according to their order $d$ and then to each category $(c_{0},c_{1},\cdots;d_{2},d_{3},\cdots)$ according to (\ref{STmonoform}). Since our ultimate purpose is to build ghost-free Lagrangians using the GST monomials, we have suppressed the categories that are ``totally reducible'', i.e., the categories in which \emph{all} the monomials can be reduced by integrations by parts in the sense that the monomials are either total derivatives or can be expressed in terms linear combinations of monomials in other categories up to total derivatives. There is a special case, the monomials of the category $(0;0,1)$ (and thus of $d=2$) are all reducible by themselves, and thus we do not include them in Tab. \ref{tab:STbasis}, although they will be used to build factorizable monomials of $d=3,4$.		
		\item Up to $d=4$, since $c_{m}$ with $m\geq 1$ and $d_{n}$ with $n\geq 4$ are all $0$'s for the irreducible categories, instead of using the cumbersome expressions $(c_{0},c_{1},\cdots;d_{2},d_{3},\cdots)$, we simply assign each monomial a set of 3 integers $(c_{0};d_{2},d_{3})$ and classify various irreducible monomials into categories  labelled by $(c_{0};d_{2},d_{3})$. This explains the second column of Tab. \ref{tab:STbasis}.
		\item All the monomials fall into two types: unfactorizable and factorizable. We then focus on the construction of the set of unfactorizable monomials, which we denote as $\mathcal{E}^{(c_{0};d_{2},d_{3})}$. The factorizable monomials can be easily got by products of the unfactorizable monomials. The unfactorizable and the factorizable monomials are listed in the third and the fourth columns of Tab. \ref{tab:STbasis}, respectively. Here and throughout this work, the symbol ``$\otimes$'' is a schematic shorthand, which  reminds us how to build factorizable monomials from the \emph{symmetrized} direct product of unfactorizable monomials\footnote{For example, we have $\{a,b\} \otimes \{x,y\} = \{ax,ay,bx,by\}$ and $\{a,b\} \otimes \{a,b\} = \{ a^2,ab,b^2 \}$, etc..}.
		\item Some of the monomials of the same order $d$ are not linearly independent. That is, some of the monomials can be \emph{algebraically} reduced by linear combinations of other monomials of the same order $d$, by making use the fact that Riemann tensor acts as the commutators of covariant derivatives as well as the (anti)symmetry of the Riemann tensor, etc..
		We suppress these algebraically reducible monomials from $\mathcal{E}^{(c_{0};d_{2},d_{3})}$.
		Clearly if several monomials are not linearly independent, there are ambiguities in choosing which monomials should be kept while others should be reduced. We simply choose the most natural and convenient ones.
	\end{itemize}
After these procedures, we choose $\mathcal{E}^{(c_{0};d_{2},d_{3})}$ such that the full set of monomials of order $d$ in Tab. \ref{tab:STbasis} form a set of linearly independent monomials, in the sense that it cannot be further reduced \emph{algebraically} and an arbitrary GST polynomial of order $d$ can be expressed in terms of the linear combination of monomials in this set. 
In this sense, we may dub the set of GST monomials in Tab. \ref{tab:STbasis} as the complete basis for the GST polynomials of order $d$.
We emphasize that the overall order $d$ is crucial, since as have been observed in Ref. \cite{Gao:2020juc} that ghost-free polynomials can arise only in the linear combinations of monomials of the same order $d$.

In the following, we list the sets of linearly independent unfactorizable monomials $\mathcal{E}^{(c_{0};d_{2},d_{3})}$ and count the number of monomials in the complete basis, i.e., the dimension of $\Sigma_{d}$.
\begin{itemize}
	\item $d=0$:
	There is a single unfactorizable monomial
	\begin{equation}
	\mathcal{E}^{\left(0;0,0\right)}\equiv\left\{ \bm{E}_{1}^{\left(0;0,0\right)}\right\} ,
	\end{equation}
	where we define the monomial $\bm{E}_{1}^{\left(0;0,0\right)}\equiv\nabla_{a}\phi\nabla^{a}\phi$.
	We thus have $\dim(\Sigma_{0}) = 1$. Here and throughout this paper, we follow the notation in \cite{Gao:2020juc} and use $\bm{E}^{(c_{0};d_{2},d_{3})}_{m}$ to denote the unfactorizable monomials.
	\item $d=1$:
	There are two unfactorizable monomials
	\begin{equation}
	\mathcal{E}^{\left(0;1,0\right)}\equiv\left\{ \bm{E}_{1}^{\left(0;1,0\right)},\bm{E}_{2}^{\left(0;1,0\right)}\right\}  ,
	\end{equation}
	where we define
		\begin{eqnarray}
		\bm{E}_{1}^{\left(0;1,0\right)} & \equiv & \frac{1}{\sigma}\square\phi,\\
		\bm{E}_{2}^{\left(0;1,0\right)} & \equiv & \frac{1}{\sigma^{3}}\nabla_{a}\phi\nabla_{b}\phi\nabla^{a}\nabla^{b}\phi,
		\end{eqnarray}
	with $\sigma\equiv\sqrt{-\nabla_{a}\phi\nabla^{a}\phi}$. 
	Thus
	\begin{equation}
		\dim (\Sigma_{1}) = 2.
	\end{equation}
	\item $d=2$:
	The sets of linearly independent unfactorizable monomials are
	\begin{eqnarray}
	\mathcal{E}^{\left(0;2,0\right)} & \equiv & \left\{ \bm{E}_{1}^{\left(0;2,0\right)},\bm{E}_{2}^{\left(0;2,0\right)}\right\} ,\\
	\mathcal{E}^{\left(1;0,0\right)} & \equiv & \left\{ \bm{E}_{1}^{\left(1;0,0\right)},\bm{E}_{2}^{\left(1;0,0\right)}\right\} ,
	\end{eqnarray}
	where 
		\begin{eqnarray}
		\bm{E}_{1}^{\left(0;2,0\right)} & \equiv & \frac{1}{\sigma^{2}}\nabla_{a}\nabla_{b}\phi\nabla^{a}\nabla^{b}\phi,\\
		\bm{E}_{2}^{\left(0;2,0\right)} & \equiv & \frac{1}{\sigma^{4}}\nabla^{a}\phi\nabla^{b}\phi\nabla_{c}\nabla_{a}\phi\nabla^{c}\nabla_{b}\phi,
		\end{eqnarray}
		and
		\begin{eqnarray}
		\bm{E}_{1}^{\left(1;0,0\right)} & \equiv & \,{}^{4}\!R,\label{E100_1}\\
		\bm{E}_{2}^{\left(1;0,0\right)} & \equiv & \frac{1}{\sigma^{2}}\,{}^{4}\!R_{ab}\nabla^{a}\phi\nabla^{b}\phi.\label{E100_2}
		\end{eqnarray}
	In addition, there are $3$ factorizable monomials. According to Tab. \ref{tab:STbasis}, these are 
		\begin{equation}
			\left(\bm{E}_{1}^{\left(0;1,0\right)}\right)^{2},\quad \bm{E}_{1}^{\left(0;1,0\right)}\bm{E}_{2}^{\left(0;1,0\right)},\quad \left(\bm{E}_{2}^{\left(0;1,0\right)}\right)^{2}.
		\end{equation}
	As a result, 
		\begin{equation}
			\dim (\Sigma_{2}) = 7.
		\end{equation}
	Note there is a special category $(0;0,1)$, of which the independent monomials are
		\begin{equation}
			\mathcal{E}^{\left( 0;0,1\right)} \equiv \left\{ \bm{E}_{1}^{\left(0;0,1\right)},\bm{E}_{3}^{\left(0;0,1\right)}\right\} , \label{calE001}
		\end{equation}
	where
		\begin{eqnarray}
		\bm{E}_{1}^{\left(0;0,1\right)} & \equiv & \frac{1}{\sigma^{2}}\nabla^{a}\phi\nabla_{a}\square\phi,\label{E001_1}\\
		\bm{E}_{3}^{\left(0;0,1\right)} & \equiv & \frac{1}{\sigma^{4}}\nabla^{a}\phi\nabla^{b}\phi\nabla^{c}\phi\nabla_{a}\nabla_{b}\nabla_{c}\phi.\label{E001_3}
		\end{eqnarray}
	As we have argued before, we do not include this category in Tab. \ref{tab:STbasis} since these monomials by themselves are reducible at the level of $d=2$. Nevertheless, they will be used to build factorizable monomials of $d=3,4$.
	\item $d=3$:
	The sets of linearly independent unfactorizable monomials are chosen to be
	\begin{eqnarray}
	\mathcal{E}^{\left(0;3,0\right)} & \equiv & \left\{ \bm{E}_{1}^{\left(0;3,0\right)},\bm{E}_{2}^{\left(0;3,0\right)}\right\} ,\\
	\mathcal{E}^{\left(0;1,1\right)} & \equiv & \left\{ \bm{E}_{1}^{\left(0;1,1\right)},\bm{E}_{3}^{\left(0;1,1\right)},\bm{E}_{5}^{\left(0;1,1\right)}\right\} , \label{calE011}\\
	\mathcal{E}^{\left(1;1,0\right)} & \equiv & \left\{ \bm{E}_{1}^{\left(1;1,0\right)},\bm{E}_{2}^{\left(1;1,0\right)},\bm{E}_{3}^{\left(1;1,0\right)}\right\} ,
	\end{eqnarray}
	where the explicit expressions for the above 8 monomials can be found in Appendix \ref{app:STxpl_d3}.
	In addition, there are $8+4+4=16$ factorizable monomials. Thus
	\begin{equation}
		\dim (\Sigma_{3}) = 24.
	\end{equation}
	At this point, we note that there are also reducible categories in $d=3$, which however only contribute to reducible monomials for $d=4$ thus can be safely suppressed. On the other hand, if we go to $d>4$, these reducible monomials should be taken into account (see Tab. I in Ref. \cite{Gao:2020juc} for details).
	\item $d=4$:
	The sets of linearly independent unfactorizable monomials are chosen to be
	\begin{eqnarray}
	\mathcal{E}^{\left(0;4,0\right)} & \equiv & \left\{ \bm{E}_{1}^{\left(0;4,0\right)},\bm{E}_{2}^{\left(0;4,0\right)}\right\} ,\\
	\mathcal{E}^{\left(0;2,1\right)} & \equiv & \left\{ \bm{E}_{1}^{\left(0;2,1\right)},\bm{E}_{3}^{\left(0;2,1\right)},\bm{E}_{5}^{\left(0;2,1\right)},\bm{E}_{7}^{\left(0;2,1\right)},\bm{E}_{8}^{\left(0;2,1\right)}\right\} ,\\
	\mathcal{E}^{\left(0;0,2\right)} & \equiv & \left\{ \bm{E}_{1}^{\left(0;0,2\right)},\bm{E}_{4}^{\left(0;0,2\right)},\bm{E}_{6}^{\left(0;0,2\right)},\bm{E}_{8}^{\left(0;0,2\right)},\bm{E}_{11}^{\left(0;0,2\right)}\right\} ,\\
	\mathcal{E}^{\left(1;2,0\right)} & \equiv & \left\{ \bm{E}_{1}^{\left(1;2,0\right)},\bm{E}_{2}^{\left(1;2,0\right)},\bm{E}_{3}^{\left(1;2,0\right)},\bm{E}_{4}^{\left(1;2,0\right)},\bm{E}_{5}^{\left(1;2,0\right)},\bm{E}_{6}^{\left(1;2,0\right)},\bm{E}_{7}^{\left(1;2,0\right)}\right\} ,\\
	\mathcal{E}^{\left(2;0,0\right)} & \equiv & \left\{ \bm{E}_{1}^{\left(2;0,0\right)},\bm{E}_{2}^{\left(2;0,0\right)},\bm{E}_{3}^{\left(2;0,0\right)},\bm{E}_{4}^{\left(2;0,0\right)},\bm{E}_{5}^{\left(2;0,0\right)},\bm{E}_{6}^{\left(2;0,0\right)}\right\} ,\\
	\mathcal{E}^{\left(1;0,1\right)} & \equiv & \left\{ \bm{E}_{2}^{\left(1;0,1\right)},\bm{E}_{5}^{\left(1;0,1\right)},\bm{E}_{7}^{\left(1;0,1\right)},\bm{E}_{8}^{\left(1;0,1\right)}\right\} ,
	\end{eqnarray}
	where the explicit expressions for the above 29 monomials are given in Appendix \ref{app:STxpl_d4}.
	In addition, there are $18+16+3+16+3+4=60$ factorizable monomials. Thus 
	\begin{equation}
		\dim (\Sigma_{4}) = 89.
	\end{equation}
\end{itemize}
From the above, it is interesting that the (beyond) Horndeski theories considered in the literature are only up to $d=3$, while the case of $d=4$ has not yet been systematically explored.

As being described before, we first exhaust all the possible GST monomials and then choose the linearly independent ones to build the complete basis.
In particular, we follow exactly the notations and the definitions for the monomials in Ref. \cite{Gao:2020juc}.
As a result, the subscripts of monomials in each set $\mathcal{E}^{(c_{0};d_{2},d_{3})}$ may not be necessarily in the arithmetic order\footnote{For example, in Eq. (\ref{calE011}), there is neither $\bm{E}^{\left(0;1,1\right)}_{2}$ nor $\bm{E}^{\left(0;1,1\right)}_{4}$ as they are not linearly independent and thus are suppressed.}.
Although one may reorder the monomials in each set, we tend not to do this in the present work. 
Since although the dimension of $\Sigma_{d}$ is fixed, the monomials we choose in the complete basis are not unique and one is free to choose other set of linearly independent monomials as the complete basis.

\subsubsection{Parity violating}

The classification of parity-violating GST monomials is completely parallel to that of the parity-preserving monomials. The results are summarized in Tab. \ref{tab:STbasis_p}.
	\begin{table}[H]
		\begin{centering}
			\begin{tabular}{|c|c|l|c|}
				\hline 
				$d$ & $\left(c_{0};d_{2},d_{3}\right)$ & Unfactorizable & Factorizable\tabularnewline
				\hline 
				\hline 
				3 & $\left(1;1,0\right)$ & $\mathcal{F}^{\left(1;1,0\right)}$ & -\tabularnewline
				\hline 
				\hline 
				4 & $\left(0;2,1\right)$ & $\mathcal{F}^{\left(0;2,1\right)}$ & -\tabularnewline
				\hline 
				& $\left(1;2,0\right)$ & $\mathcal{F}^{\left(1;2,0\right)}$ & $\mathcal{F}^{\left(1;1,0\right)}\otimes\mathcal{E}^{\left(0;1,0\right)}$\tabularnewline
				\hline 
				& $\left(2;0,0\right)$ & $\mathcal{F}^{\left(2;0,0\right)}$ & -\tabularnewline
				\hline 
				& $\left(1;0,1\right)$ & $\mathcal{F}^{\left(1;0,1\right)}$ & -\tabularnewline
				\hline 
			\end{tabular}
			\par\end{centering}
		\caption{Classification of the parity-violating GST monomials.}
		\label{tab:STbasis_p}
	\end{table}
Essentially the structure of Tab. \ref{tab:STbasis_p} is the same as that of Tab. \ref{tab:STbasis}. The main difference comes from the fact that it is not possible to build parity-violating SCG monomials in most of the categories.
In particular, there is no parity violating monomials for $d=0,1,2$ and we only list the categories of which the sets of monomials are not empty.
In the following we list the sets of unfactorizable parity-violating monomials $\mathcal{F}^{\left(c_{0};d_{2},d_{3}\right)}$.
\begin{itemize}
	\item $d=3$:
	There is only a single unfactorizable term
	\begin{equation}
	\mathcal{F}^{\left(1;1,0\right)} \equiv  \left\{ \bm{F}_{1}^{\left(1;1,0\right)}\right\} , \label{calF_110}
	\end{equation}
	where
		\begin{equation}
		\bm{F}_{1}^{\left(1;1,0\right)}\equiv\frac{1}{\sigma^{3}}\varepsilon_{abcd}\,{}^{4}\!R_{ef}^{\phantom{ef}cd}\nabla^{a}\phi\nabla^{e}\phi\nabla^{b}\nabla^{f}\phi.\label{F110_1}
		\end{equation}
	We thus have
		\begin{equation}
			\dim (\Xi_{3}) = 1.
		\end{equation}
	\item $d=4$:
	The sets of linearly independent unfactorizable monomials are chosen to be
	\begin{eqnarray}
	\mathcal{F}^{\left(0;2,1\right)} & \equiv & \left\{ \bm{F}_{6}^{\left(0;2,1\right)}\right\} ,\\
	\mathcal{F}^{\left(1;2,0\right)} & \equiv & \left\{ \bm{F}_{1}^{\left(1;2,0\right)},\bm{F}_{2}^{\left(1;2,0\right)},\bm{F}_{3}^{\left(1;2,0\right)},\bm{F}_{4}^{\left(1;2,0\right)},\bm{F}_{5}^{\left(1;2,0\right)},\bm{F}_{6}^{\left(1;2,0\right)},\bm{F}_{7}^{\left(1;2,0\right)},\bm{F}_{8}^{\left(1;2,0\right)}\right\} ,\\
	\mathcal{F}^{\left(2;0,0\right)} & \equiv & \left\{ \bm{F}_{1}^{\left(2;0,0\right)},\bm{F}_{2}^{\left(2;0,0\right)},\bm{F}_{3}^{\left(2;0,0\right)},\bm{F}_{4}^{\left(2;0,0\right)},\bm{F}_{5}^{\left(2;0,0\right)}\right\} ,\\
	\mathcal{F}^{\left(1;0,1\right)} & \equiv & \left\{ \bm{F}_{4}^{\left(1;0,1\right)}\right\} ,
	\end{eqnarray}
	where the explicit expressions for the above 15 monomials are given in Appendix \ref{app:STxpl_d4p}.
	According to Tab. \ref{tab:STbasis_p}, there are also $2$ factorizable monomials, i.e.,
	\begin{equation}
		\bm{F}_{1}^{\left(1;1,0\right)}\bm{E}_{1}^{\left(0;1,0\right)},\quad\bm{F}_{1}^{\left(1;1,0\right)}\bm{E}_{2}^{\left(0;1,0\right)}.
	\end{equation}
	Thus we have 
		\begin{equation}
			\dim (\Xi_{4}) = 17.
		\end{equation}
\end{itemize}

\subsection{SCG monomials} \label{sec:SCGmono}

The construction and classification of the SCG monomials are exactly the same as those of the GST monomials. 
By making use Eqs. (\ref{coord_1})-(\ref{coord_3}), we can also assign SCG monomial a set of integers $(c_{0};d_{2},d_{3})$ and make similar tables as Tabs. \ref{tab:STbasis} and \ref{tab:STbasis_p}.
Precisely, we denote the sets of linearly independent unfactorizable and irreducible SCG monomials as $\mathcal{X}^{(c_{0};d_{2},d_3)}$ and $\mathcal{Y}^{(c_{0};d_{2},d_{3})}$ in the parity preserving and violating cases, respectively.
These are nothing but the counterparts of $\mathcal{E}^{(c_{0};d_{2},d_{3})}$ and $\mathcal{F}^{(c_{0};d_{2},d_{3})}$ for the ST monomials. 
Moreover, the orders of various categories exactly follow Tabs. \ref{tab:STbasis} and \ref{tab:STbasis_p}, excepts a few special differences we shall describe below.

\subsubsection{Parity preserving}

In the following we summarize the parity-preserving SCG monomials up to $d=4$, and refer to Ref. \cite{Gao:2020yzr} for more details. 
Note we have suppressed all the SCG monomials that are reducible by integrations by parts from the beginning\footnote{This is different from the classification of the GST monomials, where we only suppressed the categories that are ``totally reducible'' by integrations by parts, while for monomials in the categories that are not totally reducible, we only eliminate those that are algebraically reducible.}.
\begin{itemize}
	\item $d=1$: 
	There is a single monomial
		\begin{eqnarray}
		\mathcal{X}^{\left(0;1,0\right)} = \{K\}, \label{calX_1}
		\end{eqnarray}
		which is unfactorizable and irreducible.
		Thus 
		\begin{equation}
			\dim \left(\mathcal{G}_{1}\right) = 1.
		\end{equation}
	\item $d=2$: The sets of unfactorizable and irreducible monomials are
	\begin{eqnarray}
	\mathcal{X}^{\left(0;2,0\right)} & = & \left\{ K_{ij}K^{ij},a_{i}a^{i}\right\} ,\\
	\mathcal{X}^{\left(1;0,0\right)} & = & \left\{ ^{3}\!R\right\} .
	\end{eqnarray}
	In addition, there is a single factorizable monomial $K^2$, which simply comes from $\mathcal{X}^{(0;1,0)}\otimes\mathcal{X}^{(0;1,0)}$.
	Thus
		\begin{equation}
		\dim\left(\mathcal{G}_{2}\right)=3+1=4.
		\end{equation}
	Note there is a special category\footnote{This is exactly what happens for $\mathcal{E}^{(0;0,1)}$ in Eq. (\ref{calE001}).}
	\begin{equation}
	\mathcal{X}^{\left(0;0,1\right)}=\left\{ \nabla_{i}a^{i}\right\} ,
	\end{equation}
	which is reducible at order $d=2$, but will contribute to the factorizable
	monomials of $d=3,4$. 
	\item $d=3$:
	The sets of 4 unfactorizable and irreducible monomials are
		\begin{eqnarray}
		\mathcal{X}^{\left(0;3,0\right)} & \equiv & \left\{ K_{ij}K^{jk}K_{k}^{i},K_{ij}a^{i}a^{j}\right\} ,\\
		\mathcal{X}^{\left(0;1,1\right)} & \equiv & \left\{ K_{ij}\nabla^{i}a^{j}\right\} ,\\
		\mathcal{X}^{\left(1;1,0\right)} & \equiv & \left\{ ^{3}\!R^{ij}K_{ij}\right\} .
		\end{eqnarray}
	According to Tab. \ref{tab:STbasis}, in the case of SCG monomials, there are $2+1+1+1=5$ factorizable monomials, of which the expressions can be read easily.
	We thus have
	\begin{equation}
	\dim\left(\mathcal{G}_{3}\right)=4+5=9.
	\end{equation}
	At this point, note similar to the case of ST monomials, the reducible categories of $d=3$ will also contribute to the reducible categories of $d=4$ can thus can be safely neglected in our consideration.
	\item $d=4$: 
	The sets of 12 unfactorizable and irreducible monomials are
		\begin{eqnarray}
		\mathcal{X}^{\left(0;4,0\right)} & \equiv & \left\{ K_{ik}K_{j}^{k}a^{i}a^{j}\right\} ,\\
		\mathcal{X}^{\left(0;2,1\right)} & \equiv & \left\{ K_{i}^{k}K_{jk}\nabla^{i}a^{j},K_{j}^{i}a^{j}\nabla_{k}K_{i}^{k},K_{j}^{i}a^{j}\nabla_{i}K\right\} ,\\
		\mathcal{X}^{\left(0;0,2\right)} & \equiv & \left\{ \nabla_{k}K_{ij}\nabla^{k}K^{ij},\nabla_{i}K^{ij}\nabla_{k}K_{j}^{k},\nabla_{i}K^{ij}\nabla_{j}K,\nabla_{i}K\nabla^{i}K,\nabla_{i}a_{j}\nabla^{i}a^{j}\right\} ,\\
		\mathcal{X}^{\left(1;2,0\right)} & \equiv & \left\{ ^{3}\!R_{ij}K_{k}^{i}K^{jk},{}^{3}\!R_{ij}a^{i}a^{j}\right\} ,\\
		\mathcal{X}^{\left(2;0,0\right)} & \equiv & \left\{ ^{3}\!R_{ij}\,{}^{3}\!R^{ij}\right\} .
		\end{eqnarray}
	Note there is no unfactorizable and irreducible parity-preserving SCG monomials of the category $\left(1;0,1\right)$.
	According to Tab. \ref{tab:STbasis}, there are also $8+4+1+4+1+1=19$ factorizable monomials.
	We thus have
	\begin{equation}
	\dim\left(\mathcal{G}_{4}\right)=12+19=31.
	\end{equation}
\end{itemize}

\subsubsection{Parity violating}

For the parity violating case, we denote the set of irreducible monomials as $\mathcal{Z}_{d}$, which is thus the counterpart of $\mathcal{G}_{d}$. Similar to the GST, there is no parity violating monomial of $d=0,1,2$.
	\begin{itemize}
		\item $d=3$: There is a single unfactorizable monomial
		\begin{equation}
		\mathcal{Y}^{\left(0;1,1\right)} \equiv \left\{\bm{Y}^{(0;1,1)}_{1}\right\}=  \left\{ \varepsilon_{ijk}K_{l}^{i}\nabla^{j}K^{kl} \right\} , \label{calY_011}
		\end{equation}
		which is also irreducible. Thus we have
		\begin{equation}
		\dim\left(\mathcal{Z}_{3}\right)=1.
		\end{equation}
		Note in the case of GST monomials, in Tab. \ref{tab:STbasis_p} the category is chosen to be $(1;1,0)$\footnote{This difference is completely notational. In fact, there is also a single parity-violating GST monomial $\bm{F}^{(0;1,1)}$ of the category $(0;1,1)$, which is algebraically proportional to $\bm{F}^{(1;1,0)}$. We prefer to choose $\bm{F}^{(1;1,0)}$ as the linearly independent monomial. Alternatively, one is free to choose $\bm{F}^{(0;1,1)}$ as the independent monomial.}.
		\item $d=4$: The sets of 5 unfactorizable and irreducible monomials are
			\begin{eqnarray}
			\mathcal{Y}^{\left(0;2,1\right)} & \equiv & \left\{ \varepsilon_{ijk}K^{im}K^{jn}\nabla_{m}K_{n}^{k},\varepsilon_{ijk}K^{mn}K_{m}^{i}\nabla^{j}K_{n}^{k},\varepsilon_{ijk}K_{l}^{i}a^{j}\nabla^{k}a^{l}\right\} ,\\
			\mathcal{Y}^{\left(1;2,0\right)} & \equiv & \left\{ \varepsilon_{ijk} {}^{3}\! R_{l}^{i}K^{jl}a^{k}\right\} ,\\
			\mathcal{Y}^{\left(1;0,1\right)} & \equiv & \left\{ \varepsilon_{ijk}{}^{3}\! R_{l}^{i}\nabla^{j}K^{kl}\right\} .
			\end{eqnarray}
		Note there is no irreducible parity-violating SCG monomial of  category $(2;0,0)$. In addition, this is a single factorizable monomial coming from $\mathcal{Y}^{\left(0;1,1\right)}\otimes\mathcal{X}^{\left(0;1,0\right)} = \left\{ \varepsilon_{ijk}K_{l}^{i}\nabla^{j}K^{kl}K \right\}$. Thus we have
		\begin{equation}
			\dim\left(\mathcal{Z}_{4}\right)=5+1 = 6.
		\end{equation}
	\end{itemize}

Before end this section, it is interesting to note that the dimensions of the SCG spaces  $\mathcal{G}_{d}$ and $\mathcal{Z}_{d}$ are much smaller than the dimensions of the GST spaces $\Sigma_{d}$ and $\Xi_{d}$.
This is actually what we desire, i.e., to find much smaller subspaces of the full spaces of GST and use these subspaces as our starting point to explore the ghostfree theories.

\section{$d=1,2,3$} \label{sec:d123}

In this and the next sections, we derive the GST polynomials corresponding to each SCG monomials by using the Stueckelberg trick.
In particular, the correspondences are encoded in the projection matrices $\Phi_{a\alpha}^{(d)}$ and $\Psi_{a\alpha}^{(d)}$ defined in (\ref{Phimat_def}) and (\ref{Psimat_def}) for the parity-preserving and parity-violating cases, respectively.
The GST correspondences of the SCG monomials of $d=1,2,3$ have been got in \cite{Gao:2020yzr}. Here we reformulate the results in the formalism of the current work.

\subsection{$d=1$}

The case of $d=1$ is simple, which we shall use to illustrate our formalism. From (\ref{calX_1}) there is a single SCG monomial $K$. After making use of the Stueckelberg trick, we find
	\begin{equation}
		K\rightarrow-\bm{E}_{1}^{\left(0;1,0\right)}-\bm{E}_{2}^{\left(0;1,0\right)}. \label{K_STcorr}
	\end{equation}
In our formalism, we write
	\begin{equation}
		 \mathcal{G}_{1} \equiv \left\{ \bm{g}_{1}^{(1)}\right\} =\left\{ K\right\} , 
	\end{equation}
and
	\begin{equation}
		\mathcal{E}^{\left(0;1,0\right)} \equiv \left\{ \bm{\sigma}_{\alpha}^{(1)}\right\}  =\left\{ \bm{E}_{1}^{\left(0;1,0\right)},\bm{E}_{2}^{\left(0;1,0\right)}\right\} .
	\end{equation}
The GST correspondence of $\bm{g}^{(1)}_{1}\equiv K$ is thus $\tilde{\bm{g}}_{1}^{(1)}=-\bm{E}_{1}^{\left(0;1,0\right)}-\bm{E}_{2}^{\left(0;1,0\right)}$.
Thus (\ref{K_STcorr}) is equivalent to writing
	\begin{equation}
		\tilde{\bm{g}}_{1}^{(1)}=\sum_{\alpha=1}^{2}\Phi_{1\alpha}^{(1)}\bm{\sigma}_{\alpha}^{(1)},
	\end{equation}
where $\Phi_{1\alpha}^{(1)}$ is simply 
	\begin{equation}
		\Phi_{1\alpha}^{(1)}=\left(\begin{array}{cc}
		-1 & -1\end{array}\right).
	\end{equation}

\subsection{$d=2$}

For $d=2$, since $\dim(\mathcal{G}_{2}) = 4$ and $\dim (\Sigma_{2}) = 7$, we write
	\begin{equation}
		\tilde{\bm{g}}_{a}^{(2)}=\sum_{\alpha=1}^{7}\Phi_{a\alpha}^{(2)}\bm{\sigma}_{\alpha}^{(2)},\qquad a=1,\cdots,4.
	\end{equation}
Instead of giving the expression for $\Phi_{a\alpha}^{(2)}$ directly, we split $\Phi_{a\alpha}^{(2)}$ into sub-matrices according to the categories of $(c_0;d_{2},d_{3})$.
Note this splitting is merely technical, since the matrix $\Phi^{(3)}_{a\alpha}$ and especially $\Phi^{(4)}_{a\alpha}$ become huge and unreadable.
For $d=2$, according to Tab. \ref{tab:STbasis} there are two categories of $(c_0;d_{2},d_{3})$, which we denote briefly as
\begin{equation}
\bm{1}\equiv\left(0;2,0\right),\qquad\bm{2}\equiv\left(1;0,0\right).
\end{equation}
	
After some manipulations, we find
	\begin{equation}
	\Phi_{a\alpha}^{(2)}=\left(\begin{array}{cc}
	\Phi_{\bm{1},\bm{1}}^{(2)} & \bm{0}\\
	\Phi_{\bm{2},\bm{1}}^{(2)} & \Phi_{\bm{2},\bm{2}}^{(2)}
	\end{array}\right),
	\end{equation}
which is a $4\times 7$ matrix.
The non-vanishing sub-matrices are
	\begin{equation}
	\bm{\Phi}_{\bm{1},\bm{1}}^{(2)}=\left(\begin{array}{ccccc}
	1 & 2 & 0 & 0 & 1\\
	0 & 0 & 1 & 2 & 1\\
	0 & 1 & 0 & 0 & 1
	\end{array}\right),
	\end{equation}
	\begin{equation}
	\bm{\Phi}_{\bm{2},\bm{1}}^{(2)}=\left(\begin{array}{ccccc}
	1 & 2 & -1 & -2 & 0\end{array}\right),
	\end{equation}
and
	\begin{equation}
	\bm{\Phi}_{\bm{2},\bm{2}}^{(2)}=\left(\begin{array}{cc}
	1 & 2\end{array}\right).
	\end{equation}

\subsection{$d=3$}

For $d=3$, since $\dim(\mathcal{G}_{3}) = 9$ and $\dim(\Sigma_{3})=24$, we write
	\begin{equation}
		\tilde{\bm{g}}_{a}^{(3)}=\sum_{\alpha=1}^{24}\Phi_{a\alpha}^{(3)}\bm{\sigma}_{\alpha}^{(3)},\qquad a=1,\cdots,9.
	\end{equation}
Similar to the case of $d=2$, instead of giving the expression of $\Phi_{a\alpha}^{(3)}$ directly, we split $\Phi_{a\alpha}^{(3)}$ into 3 categories according to $(c_{0};d_{2},d_{3})$, which are
	\begin{equation}
		\bm{1}\equiv\left(0;3,0\right),\quad\bm{2}\equiv\left(0;1,1\right),\quad\bm{3}\equiv\left(1;1,0\right).
	\end{equation}
We thus write
	\begin{equation}
	\Phi_{a\alpha}^{(3)}=\left(\begin{array}{ccc}
	\Phi_{\bm{1},\bm{1}}^{(3)} & \bm{0} & \bm{0}\\
	\Phi_{\bm{2},\bm{1}}^{(3)} & \Phi_{\bm{2},\bm{2}}^{(3)} & \Phi_{\bm{2},\bm{3}}^{(3)}\\
	\Phi_{\bm{3},\bm{1}}^{(3)} & \bm{0} & \Phi_{\bm{3},\bm{3}}^{(3)}
	\end{array}\right),
	\end{equation}
which is a $9\times 24$ matrix.
The non-vanishing sub-matrices are
	\begin{equation}
	\Phi_{\bm{1},\bm{1}}^{(3)}=\left(\begin{array}{cccccccccc}
	-1 & -3 & 0 & 0 & 0 & -3 & 0 & 0 & 0 & -1\\
	0 & -1 & 0 & 0 & 0 & -2 & 0 & 0 & 0 & -1\\
	0 & 0 & -1 & -2 & -1 & -2 & 0 & 0 & -1 & -1\\
	0 & 0 & 0 & -1 & 0 & -1 & 0 & 0 & -1 & -1\\
	0 & 0 & 0 & 0 & 0 & 0 & -1 & -3 & -3 & -1
	\end{array}\right),
	\end{equation}
	\begin{equation}
	\Phi_{\bm{2},\bm{1}}^{(3)}=\left(\begin{array}{cccccccccc}
	-1 & -4 & 0 & 0 & -1 & -7 & 0 & 0 & 0 & -3\\
	0 & 0 & -1 & -3 & -1 & -3 & 0 & -1 & -4 & -3
	\end{array}\right),
	\end{equation}
	\begin{equation}
	\Phi_{\bm{2},\bm{2}}^{(3)}=\left(\begin{array}{ccccccc}
	0 & -1 & -2 & 0 & 0 & 0 & -1\\
	0 & 0 & 0 & -1 & -1 & -1 & -1
	\end{array}\right),
	\end{equation}
	\begin{equation}
	\Phi_{\bm{2},\bm{3}}^{(3)}=\left(\begin{array}{ccccccc}
	0 & -1 & 0 & 0 & 0 & 0 & 0\\
	0 & 0 & 0 & 0 & 0 & -1 & -1
	\end{array}\right),
	\end{equation}
	\begin{equation}
	\Phi_{\bm{3},\bm{1}}^{(3)}=\left(\begin{array}{cccccccccc}
	-1 & -3 & 1 & 2 & 1 & -1 & 0 & 0 & 1 & 0\\
	0 & 0 & -1 & -2 & -1 & -2 & 1 & 3 & 2 & 0
	\end{array}\right),
	\end{equation}
	and
	\begin{equation}
	\Phi_{\bm{3},\bm{3}}^{(3)}=\left(\begin{array}{ccccccc}
	-1 & -1 & -2 & 0 & 0 & 0 & -1\\
	0 & 0 & 0 & -1 & -1 & -2 & -2
	\end{array}\right).
	\end{equation}

For $d=3$, there is only a single parity-violating monomials for both the GST and SCG, given in (\ref{calF_110}) and (\ref{calY_011}), respectively. The correspondence is simply
	\begin{equation}
		\bm{Y}^{\left(0;1,1\right)}_{1} \rightarrow\frac{1}{2}\bm{F}_{1}^{\left(1;1,0\right)} .
	\end{equation}

\section{$d=4$} \label{sec:d4}

In this section, we present the correspondence between the SCG and GST monomials for $d=4$, which is one of main results in this work.

\subsection{Parity preserving}

For the parity-preserving case, since $\dim(\mathcal{G}_{4}) = 31$ and $\dim(\Sigma_{4}) = 89$, we write
	\begin{equation}
		\tilde{\bm{g}}_{a}^{(4)}=\sum_{\alpha=1}^{89}\Phi_{a\alpha}^{(4)}\bm{\sigma}_{\alpha}^{(4)},\qquad a=1,\cdots,31.
	\end{equation}
According to Tab. \ref{tab:STbasis}, there are 6 categories
	\begin{eqnarray}
	\bm{1} & \equiv & \left(0;4,0\right),\quad\bm{2}\equiv\left(0;2,1\right),\quad\bm{3}\equiv\left(0;0,2\right),\\
	\bm{4} & \equiv & \left(1;2,0\right),\quad\bm{5}\equiv\left(2;0,0\right),\quad\bm{6}\equiv\left(1;0,1\right).
	\end{eqnarray}
	
We thus write
	\begin{equation}
	\Phi_{a\alpha}^{(4)}=\left(\begin{array}{cccccc}
	\Phi_{\bm{1},\bm{1}}^{(4)} & \bm{0} & \bm{0} & \bm{0} & \bm{0} & \bm{0}\\
	\Phi_{\bm{2},\bm{1}}^{(4)} & \Phi_{\bm{2},\bm{2}}^{(4)} & \bm{0} & \Phi_{\bm{2},\bm{4}}^{(4)} & \bm{0} & \bm{0}\\
	\Phi_{\bm{3},\bm{1}}^{(4)} & \Phi_{\bm{3},\bm{2}}^{(4)} & \Phi_{\bm{3},\bm{3}}^{(4)} & \Phi_{\bm{3},\bm{4}}^{(4)} & \Phi_{\bm{3},\bm{5}}^{(4)} & \Phi_{\bm{3},\bm{6}}^{(4)}\\
	\Phi_{\bm{4},\bm{1}}^{(4)} & \bm{0} & \bm{0} & \Phi_{\bm{4},\bm{4}}^{(4)} & \bm{0} & \bm{0}\\
	\Phi_{\bm{5},\bm{1}}^{(4)} & \bm{0} & \bm{0} & \Phi_{\bm{5},\bm{4}}^{(4)} & \Phi_{\bm{5},\bm{5}}^{(4)} & \bm{0}\\
	\Phi_{\bm{6},\bm{1}}^{(4)} & \Phi_{\bm{6},\bm{2}}^{(4)} & \bm{0} & \Phi_{\bm{6},\bm{4}}^{(4)} & \Phi_{\bm{6},\bm{5}}^{(4)} & \Phi_{\bm{6},\bm{6}}^{(4)}
	\end{array}\right),
	\end{equation}
which is a $31\times 89$ matrix.
After some manipulations, the non-vanishing sub-matrices are found to be
	\begin{equation}
	\Phi_{\bm{1},\bm{1}}^{(4)}=\left(\begin{array}{cccccccccccccccccccc}
	0 & 1 & 0 & 0 & 0 & 2 & 0 & 0 & 1 & 0 & 0 & 0 & 0 & 0 & 3 & 0 & 0 & 0 & 0 & 1\\
	0 & 0 & 1 & 3 & 1 & 3 & 0 & 0 & 0 & 0 & 0 & 0 & 3 & 0 & 3 & 0 & 0 & 0 & 1 & 1\\
	0 & 0 & 0 & 1 & 0 & 1 & 0 & 0 & 0 & 0 & 0 & 0 & 2 & 0 & 2 & 0 & 0 & 0 & 1 & 1\\
	0 & 0 & 0 & 0 & 0 & 0 & 1 & 4 & 4 & 0 & 0 & 0 & 0 & 2 & 4 & 0 & 0 & 0 & 0 & 1\\
	0 & 0 & 0 & 0 & 0 & 0 & 0 & 1 & 2 & 0 & 0 & 0 & 0 & 1 & 3 & 0 & 0 & 0 & 0 & 1\\
	0 & 0 & 0 & 0 & 0 & 0 & 0 & 0 & 1 & 0 & 0 & 0 & 0 & 0 & 2 & 0 & 0 & 0 & 0 & 1\\
	0 & 0 & 0 & 0 & 0 & 0 & 0 & 0 & 0 & 1 & 2 & 2 & 4 & 1 & 2 & 0 & 0 & 1 & 2 & 1\\
	0 & 0 & 0 & 0 & 0 & 0 & 0 & 0 & 0 & 0 & 1 & 0 & 2 & 0 & 1 & 0 & 0 & 1 & 2 & 1\\
	0 & 0 & 0 & 0 & 0 & 0 & 0 & 0 & 0 & 0 & 0 & 0 & 0 & 0 & 0 & 1 & 4 & 6 & 4 & 1
	\end{array}\right),
	\end{equation}
	\begin{equation}
	\Phi_{\bm{2},\bm{1}}^{(4)}=\left(\begin{array}{cccccccccccccccccccc}
	1 & 5 & 0 & 0 & 1 & 9 & 0 & 0 & 3 & 0 & 0 & 0 & 0 & 0 & 10 & 0 & 0 & 0 & 0 & 3\\
	0 & 2 & 0 & 1 & 0 & 5 & 0 & 0 & 2 & 0 & 0 & 0 & 2 & 0 & 8 & 0 & 0 & 0 & 1 & 3\\
	0 & 2 & 0 & 1 & 0 & 5 & 0 & 0 & 2 & 0 & 0 & 0 & 2 & 0 & 8 & 0 & 0 & 0 & 1 & 3\\
	0 & 0 & 1 & 4 & 1 & 4 & 0 & 0 & 0 & 0 & 0 & 1 & 7 & 1 & 7 & 0 & 0 & 0 & 3 & 3\\
	0 & 0 & 0 & 0 & 0 & 0 & 1 & 5 & 6 & 0 & 0 & 1 & 2 & 4 & 9 & 0 & 0 & 0 & 1 & 3\\
	0 & 0 & 0 & 0 & 0 & 0 & 0 & 1 & 3 & 0 & 0 & 0 & 1 & 1 & 6 & 0 & 0 & 0 & 1 & 3\\
	0 & 0 & 0 & 0 & 0 & 0 & 0 & 0 & 0 & 1 & 3 & 2 & 6 & 1 & 3 & 0 & 1 & 5 & 7 & 3
	\end{array}\right),
	\end{equation}
	\begin{equation}
	\Phi_{\bm{2},\bm{2}}^{(4)}=\left(\begin{array}{ccccccccccccccccccccc}
	0 & 0 & 1 & 2 & 1 & 0 & 0 & 0 & 0 & 0 & 2 & 0 & 0 & 0 & 1 & 0 & 0 & 0 & 0 & 0 & 1\\
	1 & 0 & 0 & 1 & 0 & 0 & 0 & 0 & 1 & 0 & 1 & 0 & 0 & 1 & 1 & 0 & 0 & 0 & 0 & 1 & 1\\
	1 & 0 & 0 & 1 & 0 & 0 & 0 & 0 & 1 & 0 & 1 & 0 & 0 & 1 & 1 & 0 & 0 & 0 & 0 & 1 & 1\\
	0 & 0 & 0 & 0 & 0 & 0 & 1 & 2 & 0 & 1 & 2 & 0 & 0 & 0 & 0 & 0 & 0 & 0 & 1 & 0 & 1\\
	0 & 0 & 0 & 0 & 0 & 0 & 0 & 0 & 0 & 0 & 0 & 1 & 1 & 2 & 2 & 0 & 0 & 0 & 0 & 1 & 1\\
	0 & 0 & 0 & 0 & 0 & 0 & 0 & 0 & 0 & 0 & 0 & 0 & 0 & 1 & 1 & 0 & 0 & 0 & 0 & 1 & 1\\
	0 & 0 & 0 & 0 & 0 & 0 & 0 & 0 & 0 & 0 & 0 & 0 & 0 & 0 & 0 & 1 & 1 & 2 & 2 & 1 & 1
	\end{array}\right),
	\end{equation}
	\begin{equation}
	\Phi_{\bm{2},\bm{4}}^{(4)}=\left(\begin{array}{ccccccccccccccccccccccc}
	0 & 0 & 1 & 0 & 0 & 0 & 1 & 0 & 0 & 0 & 0 & 0 & 0 & 0 & 0 & 0 & 0 & 0 & 0 & 0 & 0 & 0 & 0\\
	0 & 0 & 0 & 0 & 0 & 1 & 0 & 0 & 0 & 0 & 0 & 0 & 1 & 0 & 0 & 0 & 1 & 0 & 0 & 0 & 0 & 0 & 1\\
	0 & 0 & 0 & 0 & 0 & 0 & 0 & 0 & 0 & 0 & 0 & 0 & 0 & 0 & 0 & 0 & 0 & 0 & 0 & 0 & 0 & 0 & 0\\
	0 & 0 & 0 & 0 & 0 & 0 & 0 & 0 & 0 & 1 & 1 & 0 & 0 & 0 & 0 & 0 & 0 & 0 & 0 & 0 & 0 & 0 & 0\\
	0 & 0 & 0 & 0 & 0 & 0 & 0 & 0 & 0 & 0 & 0 & 0 & 0 & 0 & 0 & 1 & 2 & 0 & 0 & 0 & 0 & 0 & 1\\
	0 & 0 & 0 & 0 & 0 & 0 & 0 & 0 & 0 & 0 & 0 & 0 & 0 & 0 & 0 & 0 & 1 & 0 & 0 & 0 & 0 & 0 & 1\\
	0 & 0 & 0 & 0 & 0 & 0 & 0 & 0 & 0 & 0 & 0 & 0 & 0 & 0 & 0 & 0 & 0 & 0 & 0 & 0 & 1 & 2 & 1
	\end{array}\right),
	\end{equation}
	\begin{equation}
	\Phi_{\bm{3},\bm{1}}^{(4)}=\left(\begin{array}{cccccccccccccccccccc}
	0 & 6 & 0 & 0 & 0 & 12 & 0 & 3 & 12 & 0 & 0 & 0 & 0 & 3 & 27 & 0 & 0 & 0 & 0 & 9\\
	0 & 4 & 0 & 4 & 0 & 12 & 0 & 0 & 4 & 0 & 1 & 0 & 10 & 0 & 21 & 0 & 0 & 1 & 6 & 9\\
	0 & 4 & 0 & 4 & 0 & 12 & 0 & 0 & 4 & 0 & 1 & 0 & 10 & 0 & 21 & 0 & 0 & 1 & 6 & 9\\
	0 & 4 & 0 & 4 & 0 & 12 & 0 & 0 & 4 & 0 & 1 & 0 & 10 & 0 & 21 & 0 & 0 & 1 & 6 & 9\\
	1 & 6 & 0 & 0 & 2 & 16 & 0 & 0 & 5 & 0 & 0 & 0 & 0 & 1 & 24 & 0 & 0 & 0 & 0 & 9\\
	0 & 0 & 0 & 0 & 0 & 0 & 1 & 6 & 9 & 0 & 0 & 2 & 6 & 6 & 18 & 0 & 0 & 1 & 6 & 9
	\end{array}\right),
	\end{equation}
	\begin{equation}
	\Phi_{\bm{3},\bm{2}}^{(4)}=\left(\begin{array}{ccccccccccccccccccccc}
	0 & 6 & 0 & 0 & 12 & 0 & 0 & 0 & 0 & 6 & 18 & 0 & 0 & 0 & 0 & 0 & 0 & 0 & 0 & 0 & 6\\
	4 & 0 & 0 & 4 & 0 & 2 & 0 & 2 & 6 & 0 & 6 & 0 & 0 & 4 & 4 & 0 & 0 & 2 & 2 & 6 & 6\\
	4 & 0 & 0 & 4 & 0 & 2 & 0 & 2 & 6 & 0 & 6 & 0 & 0 & 4 & 4 & 0 & 0 & 2 & 2 & 6 & 6\\
	4 & 0 & 0 & 4 & 0 & 2 & 0 & 2 & 6 & 0 & 6 & 0 & 0 & 4 & 4 & 0 & 0 & 2 & 2 & 6 & 6\\
	0 & 0 & 2 & 4 & 4 & 0 & 0 & 0 & 0 & 2 & 12 & 0 & 0 & 0 & 2 & 0 & 0 & 0 & 0 & 0 & 6\\
	0 & 0 & 0 & 0 & 0 & 0 & 0 & 0 & 0 & 0 & 0 & 2 & 2 & 6 & 6 & 0 & 0 & 2 & 2 & 6 & 6
	\end{array}\right),
	\end{equation}
	\begin{equation}
	\Phi_{\bm{3},\bm{3}}^{(4)}=\left(\begin{array}{cccccccc}
	0 & 1 & 0 & 3 & 3 & 0 & 0 & 1\\
	1 & 0 & 2 & 0 & 1 & 1 & 2 & 1\\
	1 & 0 & 2 & 0 & 1 & 1 & 2 & 1\\
	1 & 0 & 2 & 0 & 1 & 1 & 2 & 1\\
	0 & 0 & 0 & 1 & 2 & 0 & 0 & 1\\
	0 & 0 & 0 & 0 & 0 & 1 & 2 & 1
	\end{array}\right),
	\end{equation}
	\begin{equation}
	\Phi_{\bm{3},\bm{4}}^{(4)}=\left(\begin{array}{ccccccccccccccccccccccc}
	0 & 0 & 0 & 4 & 0 & 0 & 8 & 0 & 0 & 0 & 4 & 0 & 0 & 0 & 0 & 0 & 0 & 0 & 0 & 0 & 0 & 0 & 0\\
	0 & 0 & 0 & 0 & 0 & 4 & 0 & 0 & 0 & 0 & 0 & 2 & 6 & 0 & 0 & 0 & 4 & 0 & 0 & 0 & 0 & 2 & 6\\
	0 & 0 & 0 & 0 & 0 & 2 & 0 & 0 & 0 & 0 & 0 & 1 & 3 & 0 & 0 & 0 & 2 & 0 & 0 & 0 & 0 & 1 & 3\\
	0 & 0 & 0 & 0 & 0 & 0 & 0 & 0 & 0 & 0 & 0 & 0 & 0 & 0 & 0 & 0 & 0 & 0 & 0 & 0 & 0 & 0 & 0\\
	0 & 0 & 2 & 0 & 0 & 0 & 4 & 0 & 0 & 0 & 2 & 0 & 0 & 0 & 0 & 0 & 0 & 0 & 0 & 0 & 0 & 0 & 0\\
	0 & 0 & 0 & 0 & 0 & 0 & 0 & 0 & 0 & 0 & 0 & 0 & 0 & 0 & 0 & 2 & 6 & 0 & 0 & 0 & 0 & 2 & 6
	\end{array}\right),
	\end{equation}
	\begin{equation}
	\Phi_{\bm{3},\bm{5}}^{(4)}=\left(\begin{array}{ccccccccc}
	0 & 0 & 0 & 0 & 0 & 1 & 0 & 0 & 0\\
	0 & 0 & 0 & 0 & 1 & 0 & 0 & 0 & 1\\
	0 & 0 & 0 & 0 & 0 & 0 & 0 & 0 & 0\\
	0 & 0 & 0 & 0 & 0 & 0 & 0 & 0 & 0\\
	0 & 0 & 0 & 0 & 0 & 0 & 0 & 0 & 0\\
	0 & 0 & 0 & 0 & 0 & 0 & 0 & 0 & 1
	\end{array}\right),
	\end{equation}
	\begin{equation}
	\Phi_{\bm{3},\bm{6}}^{(4)}=\left(\begin{array}{cccccccc}
	0 & 0 & -2 & 0 & 0 & 0 & 0 & 0\\
	2 & 0 & 0 & 2 & 0 & 0 & 2 & 2\\
	1 & 0 & 0 & 1 & 0 & 0 & 1 & 1\\
	0 & 0 & 0 & 0 & 0 & 0 & 0 & 0\\
	0 & 0 & 0 & 0 & 0 & 0 & 0 & 0\\
	0 & 0 & 0 & 0 & 0 & 0 & 2 & 2
	\end{array}\right),
	\end{equation}
	\begin{equation}
	\Phi_{\bm{4},\bm{1}}^{(4)}=\left(\begin{array}{cccccccccccccccccccc}
	1 & 4 & -1 & -3 & -1 & 1 & 0 & 0 & 2 & 0 & 0 & 0 & -3 & 0 & 1 & 0 & 0 & 0 & -1 & 0\\
	0 & 1 & 0 & -1 & 0 & 1 & 0 & 0 & 1 & 0 & 0 & 0 & -2 & 0 & 1 & 0 & 0 & 0 & -1 & 0\\
	0 & 0 & 1 & 3 & 1 & 3 & 0 & 0 & 0 & -1 & -2 & -2 & -1 & -1 & 1 & 0 & 0 & -1 & -1 & 0\\
	0 & 0 & 0 & 0 & 0 & 0 & 1 & 4 & 4 & -1 & -2 & -2 & -4 & 1 & 2 & 0 & 0 & -1 & -2 & 0\\
	0 & 0 & 0 & 0 & 0 & 0 & 0 & 1 & 2 & 0 & -1 & 0 & -2 & 1 & 2 & 0 & 0 & -1 & -2 & 0\\
	0 & 0 & 0 & 0 & 0 & 0 & 0 & 0 & 0 & 1 & 2 & 2 & 4 & 1 & 2 & -1 & -4 & -5 & -2 & 0
	\end{array}\right),
	\end{equation}
	\begin{equation}
	\Phi_{\bm{4},\bm{4}}^{(4)}=\left(\begin{array}{ccccccccccccccccccccccc}
	0 & 1 & 1 & 0 & 1 & 2 & 1 & 0 & 0 & 0 & 0 & 0 & 2 & 0 & 0 & 0 & 1 & 0 & 0 & 0 & 0 & 0 & 1\\
	0 & 0 & 0 & 0 & 1 & 0 & 1 & 0 & 0 & 0 & 0 & 0 & 2 & 0 & 0 & 0 & 0 & 0 & 0 & 0 & 0 & 0 & 1\\
	0 & 0 & 0 & 0 & 0 & 0 & 0 & 1 & 1 & 1 & 1 & 2 & 2 & 0 & 0 & 0 & 0 & 0 & 0 & 0 & 0 & 1 & 1\\
	0 & 0 & 0 & 0 & 0 & 0 & 0 & 0 & 0 & 0 & 0 & 0 & 0 & 1 & 2 & 2 & 4 & 0 & 0 & 1 & 0 & 0 & 2\\
	0 & 0 & 0 & 0 & 0 & 0 & 0 & 0 & 0 & 0 & 0 & 0 & 0 & 0 & 1 & 0 & 2 & 0 & 0 & 1 & 0 & 0 & 2\\
	0 & 0 & 0 & 0 & 0 & 0 & 0 & 0 & 0 & 0 & 0 & 0 & 0 & 0 & 0 & 0 & 0 & 1 & 2 & 1 & 2 & 4 & 2
	\end{array}\right),
	\end{equation}
	\begin{equation}
	\Phi_{\bm{5},\bm{1}}^{(4)}=\left(\begin{array}{cccccccccccccccccccc}
	1 & 4 & -2 & -6 & -2 & -2 & 0 & 0 & 2 & 1 & 2 & 2 & -2 & 1 & 0 & 0 & 0 & 1 & 0 & 0\\
	0 & 0 & 0 & 0 & 0 & 0 & 1 & 4 & 4 & -2 & -4 & -4 & -8 & 0 & 0 & 1 & 4 & 4 & 0 & 0
	\end{array}\right),
	\end{equation}
	\begin{equation}
	\Phi_{\bm{5},\bm{4}}^{(4)}=\left(\begin{array}{ccccccccccccccccccccccc}
	0 & 2 & 2 & 0 & 2 & 4 & 2 & -2 & -2 & -2 & -2 & -4 & 0 & 0 & 0 & 0 & 2 & 0 & 0 & 0 & 0 & -2 & 0\\
	0 & 0 & 0 & 0 & 0 & 0 & 0 & 0 & 0 & 0 & 0 & 0 & 0 & 2 & 4 & 4 & 8 & -2 & -4 & 0 & -4 & -8 & 0
	\end{array}\right),
	\end{equation}
	\begin{equation}
	\Phi_{\bm{5},\bm{5}}^{(4)}=\left(\begin{array}{ccccccccc}
	0 & 1 & 0 & 2 & 2 & 1 & 0 & 0 & 1\\
	0 & 0 & 0 & 0 & 0 & 0 & 1 & 4 & 4
	\end{array}\right),
	\end{equation}
	\begin{equation}
	\Phi_{\bm{6},\bm{1}}^{(4)}=\left(\begin{array}{cccccccccccccccccccc}
	0 & 0 & 0 & 0 & 0 & 0 & 1 & 5 & 6 & -1 & -3 & -1 & -4 & 3 & 6 & 0 & -1 & -5 & -6 & 0\end{array}\right),
	\end{equation}
	\begin{equation}
	\Phi_{\bm{6},\bm{2}}^{(4)}=\left(\begin{array}{ccccccccccccccccccccc}
	0 & 0 & 0 & 0 & 0 & 0 & 0 & 0 & 0 & 0 & 0 & 1 & 1 & 2 & 2 & -1 & -1 & -2 & -2 & 0 & 0\end{array}\right),
	\end{equation}
	\begin{equation}
	\Phi_{\bm{6},\bm{4}}^{(4)}=\left(\begin{array}{ccccccccccccccccccccccc}
	0 & 0 & 0 & 0 & 0 & 0 & 0 & 0 & 0 & 0 & 0 & 0 & 0 & 1 & 3 & 3 & 8 & 0 & 1 & 3 & -1 & 0 & 6\end{array}\right),
	\end{equation}
	\begin{equation}
	\Phi_{\bm{6},\bm{5}}^{(4)}=\left(\begin{array}{ccccccccc}
	0 & 0 & 0 & 0 & 0 & 0 & 0 & 1 & 2\end{array}\right),
	\end{equation}
and
	\begin{equation}
	\Phi_{\bm{6},\bm{6}}^{(4)}=\left(\begin{array}{cccccccc}
	0 & 0 & 0 & 0 & 1 & 1 & 2 & 2\end{array}\right).
	\end{equation}

\subsection{Parity violating}

For the parity-violating case, since $\dim(\mathcal{Z}_{4}) = 6$ and $\dim (\Xi_{4}) = 17$, we write
	\begin{equation}
		\tilde{\bm{z}}_{a}^{(4)}=\sum_{\alpha=1}^{17}\Psi_{a\alpha}^{(4)}\bm{\xi}_{\alpha}^{(4)},\qquad a=1,\cdots6.
	\end{equation}
There are 4 categories
	\begin{equation}
		\bm{1}\equiv\left(0;2,1\right),\quad\bm{2}\equiv\left(1;2,0\right),\quad\bm{3}\equiv\left(2;0,0\right),\quad\bm{4}\equiv\left(1;0,1\right).
	\end{equation}

Recall that there is no parity-violating SCG monomial of the category $\bm{3}\equiv\left(2;0,0\right)$, we thus write
	\begin{equation}
	\Psi_{a\alpha}^{(4)}=\left(\begin{array}{cccc}
	\Psi_{\bm{1},\bm{1}}^{(4)} & \Psi_{\bm{1},\bm{2}}^{(4)} & \bm{0} & \bm{0}\\
	\bm{0} & \Psi_{\bm{2},\bm{2}}^{(4)} & \bm{0} & \bm{0}\\
	\bm{0} & \Psi_{\bm{4},\bm{2}}^{(4)} & \Psi_{\bm{4},\bm{3}}^{(4)} & \bm{0}
	\end{array}\right),
	\end{equation}
which is a $6\times 17$ matrix.
The non-vanishing sub-matrices are
	\begin{equation}
	\Psi_{\bm{1},\bm{1}}^{(4)}=\left(\begin{array}{c}
	0\\
	0\\
	-1\\
	0
	\end{array}\right),
	\end{equation}
	\begin{equation}
	\Psi_{\bm{1},\bm{2}}^{(4)}=\left(\begin{array}{cccccccccc}
	0 & 0 & 0 & 0 & 1 & 0 & 0 & -1 & 0 & 0\\
	0 & -\frac{1}{2} & 0 & 0 & 0 & 0 & -\frac{1}{2} & 0 & 0 & 0\\
	0 & 0 & 0 & 0 & 0 & 0 & 0 & 0 & 0 & 0\\
	0 & 0 & 0 & 0 & 0 & 0 & 0 & 0 & -\frac{1}{2} & -\frac{1}{2}
	\end{array}\right),
	\end{equation}
	\begin{equation}
	\Psi_{\bm{2},\bm{2}}^{(4)}=\left(\begin{array}{cccccccccc}
	0 & 0 & 0 & 0 & 0 & -1 & 0 & -1 & 0 & 0\end{array}\right),
	\end{equation}
	\begin{equation}
	\Psi_{\bm{4},\bm{2}}^{(4)}=\left(\begin{array}{cccccccccc}
	0 & -\frac{1}{2} & 0 & 0 & 0 & 0 & -\frac{1}{2} & 0 & \frac{1}{2} & \frac{1}{2}\end{array}\right),
	\end{equation}
and
	\begin{equation}
	\Psi_{\bm{4},\bm{3}}^{(4)}=\left(\begin{array}{ccccc}
	0 & 0 & 0 & -\frac{1}{2} & -\frac{1}{2}\end{array}\right).
	\end{equation}

\section{Conclusion} \label{sec:con}

The main theoretical achievement of the scalar-tensor theory in the last decade is the rediscovery of the Horndeski theory and the construction of degenerate higher-order derivative scalar-tensor theory. These theories include derivatives of the scalar field up to the second order and curvature tensor up to the linear order.

This work is one of a series of attempts towards the ``next generation'' of theories, precisely, ghostfree theories with derivatives of the scalar field up to the cubic order and with the curvature tensor up to the quadratic order. 
The idea is to use the spatially covariant gravity (SCG) to generate ghostfree higher derivative scalar-tensor theories (GST).
Following \cite{Gao:2020juc,Gao:2020yzr}, we make a general linear algebraic analysis in Sec. \ref{sec:ls}. The various linear spaces and maps are summarized in Fig. \ref{fig:maps}.
We systematically exhaust and classify the monomials of both the GST and the SCG in Sec. \ref{sec:STmono} and Sec. \ref{sec:SCGmono}, respectively.
The final results are the ``complete basis'' for the corresponding monomials.
For the GST monomials, the results are summarized in Tab. \ref{tab:STbasis} for the parity preserving case, and in Tab. \ref{tab:STbasis_p} for the parity violating case.
A similar classification can be made for the SCG monomials.

Since the gauge fixing/recovering mappings between the GST and SCG terms are one-to-one, there is a well-defined GST subspace $\Gamma_{d}$ for each $d$, where $d$ is the total number of derivatives in the framework of SCG.
$\Gamma_{d}$ is the image of the SCG subspace $\mathcal{G}_{d}$, which is built of linearly independent SCG monomials containing only spatial derivatives and thus propagates at most 3 DoFs.
Each basis of the SCG subspace $\mathcal{G}_{d}$ (i.e., a SCG monomial) is mapped to a ``vector'' in the GST space (i.e., a GST polynomial), which is automatically ghostfree as long as the scalar field is timelike.
The main task in this work is to derive all the expressions for these image vectors in terms of the GST complete basis.
In particular, we may view the subspace $\Gamma_{d}$ as being projected from the original space $\Sigma_{d}$.
We derive the explicit expressions for the projection matrices in Sec. \ref{sec:d123} for $d=1,2,3$ and in Sec. \ref{sec:d4} for $d=4$.

The linear algebraic structure revealed in this work may be useful in exploring the subspace of scalar-tensor theory that is ghostfree ``absolutely'', i.e., irrelevant to the configuration of the scalar field. We shall investigate this in future publications.

\acknowledgments

This work was partly supported by the Natural Science Foundation of China (NSFC) under the grant No. 11975020.

\appendix

\section{Explicit expressions for the GST monomials}

In this appendix we show the explicit expressions for the linearly independent unfactorizable GST monomials, which are chosen to be in the complete basis. 
The purpose is for the completeness and self-contained.
A full list of the expressions of all unfactorizable monomials and their linear dependence can be found in Ref. \cite{Gao:2020juc}.

\subsection{Parity preserving}

\subsubsection{$d=3$} \label{app:STxpl_d3}

We define
	\begin{eqnarray}
	\bm{E}_{1}^{\left(0;3,0\right)} & \equiv & \frac{1}{\sigma^{3}}\nabla_{a}\nabla^{b}\phi\nabla_{b}\nabla^{c}\phi\nabla_{c}\nabla^{a}\phi,\\
	\bm{E}_{2}^{\left(0;3,0\right)} & \equiv & \frac{1}{\sigma^{5}}\nabla^{a}\phi\nabla^{b}\phi\nabla_{a}\nabla_{c}\phi\nabla^{c}\nabla_{d}\phi\nabla^{d}\nabla_{b}\phi,
	\end{eqnarray}
	\begin{eqnarray}
	\bm{E}_{1}^{\left(0;1,1\right)} & \equiv & \frac{1}{\sigma^{3}}\nabla^{a}\phi\nabla_{a}\nabla^{b}\phi\nabla_{b}\square\phi,\\
	\bm{E}_{3}^{\left(0;1,1\right)} & \equiv & \frac{1}{\sigma^{3}}\nabla^{a}\phi\nabla^{b}\nabla^{c}\phi\nabla_{a}\nabla_{b}\nabla_{c}\phi,\\
	\bm{E}_{5}^{\left(0;1,1\right)} & \equiv & \frac{1}{\sigma^{5}}\nabla^{a}\phi\nabla^{b}\phi\nabla^{c}\phi\nabla_{a}\nabla^{d}\phi\nabla_{d}\nabla_{b}\nabla_{c}\phi,\label{E011_5}
	\end{eqnarray}
and
	\begin{eqnarray}
	\bm{E}_{1}^{\left(1;1,0\right)} & \equiv & \frac{1}{\sigma}\,{}^{4}\!R_{ab}\nabla^{a}\nabla^{b}\phi,\\
	\bm{E}_{2}^{\left(1;1,0\right)} & \equiv & \frac{1}{\sigma^{3}}\,{}^{4}\!R_{abcd}\nabla^{a}\phi\nabla^{c}\phi\nabla^{b}\nabla^{d}\phi,\\
	\bm{E}_{3}^{\left(1;1,0\right)} & \equiv & \frac{1}{\sigma^{3}}\,{}^{4}\!R_{ab}\nabla^{a}\phi\nabla^{c}\phi\nabla^{b}\nabla_{c}\phi.
	\end{eqnarray}

\subsubsection{$d=4$} \label{app:STxpl_d4}

We define
	\begin{eqnarray}
	\bm{E}_{1}^{\left(0;4,0\right)} & \equiv & \frac{1}{\sigma^{4}}\nabla^{a}\nabla^{b}\phi\nabla^{c}\nabla_{b}\phi\nabla_{c}\nabla_{d}\phi\nabla^{d}\nabla_{a}\phi,\\
	\bm{E}_{2}^{\left(0;4,0\right)} & \equiv & \frac{1}{\sigma^{6}}\nabla^{a}\phi\nabla^{b}\phi\nabla_{a}\nabla_{c}\phi\nabla^{c}\nabla_{d}\phi\nabla^{d}\nabla^{e}\phi\nabla_{e}\nabla_{b}\phi,
	\end{eqnarray}
	\begin{eqnarray}
	\bm{E}_{1}^{\left(0;2,1\right)} & \equiv & \frac{1}{\sigma^{4}}\nabla^{a}\phi\nabla_{a}\nabla^{b}\phi\nabla^{c}\nabla_{b}\phi\nabla_{c}\square\phi,\\
	\bm{E}_{3}^{\left(0;2,1\right)} & \equiv & \frac{1}{\sigma^{4}}\nabla^{a}\phi\nabla_{a}\nabla^{b}\phi\nabla^{c}\nabla^{d}\phi\nabla_{b}\nabla_{c}\nabla_{d}\phi,\\
	\bm{E}_{5}^{\left(0;2,1\right)} & \equiv & \frac{1}{\sigma^{4}}\nabla^{a}\phi\nabla^{b}\nabla^{c}\phi\nabla^{d}\nabla_{c}\phi\nabla_{a}\nabla_{b}\nabla_{d}\phi,\\
	\bm{E}_{7}^{\left(0;2,1\right)} & \equiv & \frac{1}{\sigma^{6}}\nabla^{a}\phi\nabla^{b}\phi\nabla^{c}\phi\nabla^{d}\nabla_{a}\phi\nabla^{e}\nabla_{d}\phi\nabla_{e}\nabla_{b}\nabla_{c}\phi,\\
	\bm{E}_{8}^{\left(0;2,1\right)} & \equiv & \frac{1}{\sigma^{6}}\nabla^{a}\phi\nabla^{b}\phi\nabla^{c}\phi\nabla^{d}\nabla_{a}\phi\nabla^{e}\nabla_{b}\phi\nabla_{c}\nabla_{d}\nabla_{e}\phi,
	\end{eqnarray}
	\begin{eqnarray}
	\bm{E}_{1}^{\left(0;0,2\right)} & \equiv & \frac{1}{\sigma^{2}}\nabla_{a}\square\phi\nabla^{a}\square\phi,\\
	\bm{E}_{4}^{\left(0;0,2\right)} & \equiv & \frac{1}{\sigma^{2}}\nabla_{a}\nabla_{b}\nabla_{c}\phi\nabla^{a}\nabla^{b}\nabla^{c}\phi,\\
	\bm{E}_{6}^{\left(0;0,2\right)} & \equiv & \frac{1}{\sigma^{4}}\nabla^{a}\phi\nabla^{b}\phi\nabla^{c}\nabla_{a}\nabla_{b}\phi\nabla_{c}\square\phi,\\
	\bm{E}_{8}^{\left(0;0,2\right)} & \equiv & \frac{1}{\sigma^{4}}\nabla^{a}\phi\nabla^{b}\phi\nabla_{c}\nabla_{d}\nabla_{a}\phi\nabla^{c}\nabla^{d}\nabla_{b}\phi,\\
	\bm{E}_{11}^{\left(0;0,2\right)} & \equiv & \frac{1}{\sigma^{6}}\nabla^{a}\phi\nabla^{b}\phi\nabla^{c}\phi\nabla^{d}\phi\nabla^{e}\nabla_{a}\nabla_{b}\phi\nabla_{e}\nabla_{c}\nabla_{d}\phi,
	\end{eqnarray}
	\begin{eqnarray}
	\bm{E}_{1}^{\left(1;2,0\right)} & \equiv & \frac{1}{\sigma^{2}}\,{}^{4}\!R_{abcd}\,\nabla^{a}\nabla^{c}\phi\nabla^{b}\nabla^{d}\phi,\\
	\bm{E}_{2}^{\left(1;2,0\right)} & \equiv & \frac{1}{\sigma^{2}}\,{}^{4}\!R^{ab}\,\nabla_{a}\nabla^{c}\phi\nabla_{b}\nabla_{c}\phi,\\
	\bm{E}_{3}^{\left(1;2,0\right)} & \equiv & \frac{1}{\sigma^{4}}\,{}^{4}\!R_{abcd}\nabla^{a}\phi\nabla^{c}\phi\nabla^{b}\nabla^{e}\phi\nabla^{d}\nabla_{e}\phi,\\
	\bm{E}_{4}^{\left(1;2,0\right)} & \equiv & \frac{1}{\sigma^{4}}\,{}^{4}\!R_{abcd}\nabla^{a}\phi\nabla^{e}\phi\nabla^{c}\nabla_{e}\phi\nabla^{b}\nabla^{d}\phi,\\
	\bm{E}_{5}^{\left(1;2,0\right)} & \equiv & \frac{1}{\sigma^{4}}\,{}^{4}\!R^{ab}\nabla^{c}\phi\nabla^{d}\phi\nabla_{a}\nabla_{c}\phi\nabla_{b}\nabla_{d}\phi,\\
	\bm{E}_{6}^{\left(1;2,0\right)} & \equiv & \frac{1}{\sigma^{4}}\,{}^{4}\!R^{ab}\nabla_{a}\phi\nabla^{c}\phi\nabla_{b}\nabla_{d}\phi\nabla_{c}\nabla^{d}\phi,\\
	\bm{E}_{7}^{\left(1;2,0\right)} & \equiv & \frac{1}{\sigma^{6}}\,{}^{4}\!R_{abcd}\nabla^{a}\phi\nabla^{c}\phi\nabla^{f}\phi\nabla^{e}\phi\nabla^{b}\nabla_{f}\phi\nabla^{d}\nabla_{e}\phi,
	\end{eqnarray}
	\begin{eqnarray}
	\bm{E}_{1}^{\left(2;0,0\right)} & \equiv & \,{}^{4}\!R_{abcd}\,{}^{4}\!R^{abcd},\\
	\bm{E}_{2}^{\left(2;0,0\right)} & \equiv & \,{}^{4}\!R_{ab}\,{}^{4}\!R^{ab},\\
	\bm{E}_{3}^{\left(2;0,0\right)} & \equiv & \frac{1}{\sigma^{2}}\,{}^{4}\!R_{a}^{\phantom{a}cde}\,{}^{4}\!R_{bcde}\,\nabla^{a}\phi\nabla^{b}\phi,\\
	\bm{E}_{4}^{\left(2;0,0\right)} & \equiv & \frac{1}{\sigma^{2}}\,{}^{4}\!R_{acbd}\,{}^{4}\!R^{ab}\,\nabla^{c}\phi\nabla^{d}\phi,\\
	\bm{E}_{5}^{\left(2;0,0\right)} & \equiv & \frac{1}{\sigma^{2}}\,{}^{4}\!R_{ac}\,{}^{4}\!R_{\phantom{c}b}^{c}\,\nabla^{a}\phi\nabla^{b}\phi,\\
	\bm{E}_{6}^{\left(2;0,0\right)} & \equiv & \frac{1}{\sigma^{4}}\,{}^{4}\!R_{a\phantom{e}b}^{\phantom{a}e\phantom{b}f}\,{}^{4}\!R_{cedf}\nabla^{a}\phi\nabla^{b}\phi\nabla^{c}\phi\nabla^{d}\phi,
	\end{eqnarray}
and
	\begin{eqnarray}
	\bm{E}_{2}^{\left(1;0,1\right)} & \equiv & \frac{1}{\sigma^{2}}\,{}^{4}\!R^{ab}\nabla_{a}\phi\nabla_{b}\square\phi,\\
	\bm{E}_{5}^{\left(1;0,1\right)} & \equiv & \frac{1}{\sigma^{2}}\,{}^{4}\!R^{ab}\nabla^{c}\phi\nabla_{a}\nabla_{b}\nabla_{c}\phi,\\
	\bm{E}_{7}^{\left(1;0,1\right)} & \equiv & \frac{1}{\sigma^{4}}\,{}^{4}\!R_{abcd}\nabla^{a}\phi\nabla^{c}\phi\nabla^{e}\phi\nabla^{b}\nabla^{d}\nabla_{e}\phi,\\
	\bm{E}_{8}^{\left(1;0,1\right)} & \equiv & \frac{1}{\sigma^{4}}\,{}^{4}\!R^{ab}\nabla_{a}\phi\nabla^{c}\phi\nabla^{d}\phi\nabla_{b}\nabla_{c}\nabla_{d}\phi.
	\end{eqnarray}

\subsection{Parity violating}

\subsubsection{$d=4$} \label{app:STxpl_d4p}

We define
	\begin{equation}
	\bm{F}_{6}^{\left(0;2,1\right)}\equiv\frac{1}{\sigma^{6}}\varepsilon_{abcd}\nabla^{e}\phi\nabla^{f}\phi\nabla^{a}\phi\nabla^{b}\nabla_{e}\phi\nabla^{c}\nabla^{m}\phi\nabla_{m}\nabla_{f}\nabla^{d}\phi,\label{F021_6}
	\end{equation}
	\begin{eqnarray}
	\bm{F}_{1}^{\left(1;2,0\right)} & \equiv & \frac{1}{\sigma^{2}}\varepsilon_{abcd}\,{}^{4}\!R_{ef}^{\phantom{ef}cd}\nabla^{a}\nabla^{e}\phi\nabla^{b}\nabla^{f}\phi,\\
	\bm{F}_{2}^{\left(1;2,0\right)} & \equiv & \frac{1}{\sigma^{4}}\varepsilon_{abcd}\,{}^{4}\!R_{ef}^{\phantom{ef}cd}\nabla^{a}\phi\nabla^{e}\phi\nabla^{b}\nabla_{m}\phi\nabla^{f}\nabla^{m}\phi,\\
	\bm{F}_{3}^{\left(1;2,0\right)} & \equiv & \frac{1}{\sigma^{4}}\varepsilon_{abcd}\,{}^{4}\!R_{ef}^{\phantom{ef}cd}\nabla^{e}\phi\nabla^{m}\phi\nabla^{a}\nabla_{m}\phi\nabla^{b}\nabla^{f}\phi,\\
	\bm{F}_{4}^{\left(1;2,0\right)} & \equiv & \frac{1}{\sigma^{4}}\varepsilon_{abcd}\,{}^{4}\!R_{ef}^{\phantom{ef}cd}\nabla^{a}\phi\nabla^{m}\phi\nabla^{b}\nabla^{e}\phi\nabla^{f}\nabla_{m}\phi,\\
	\bm{F}_{5}^{\left(1;2,0\right)} & \equiv & \frac{1}{\sigma^{4}}\varepsilon_{abcd}\,{}^{4}\!R_{ef}^{\phantom{ef}cm}\nabla^{a}\phi\nabla^{e}\phi\nabla^{b}\nabla^{f}\phi\nabla^{d}\nabla_{m}\phi,\\
	\bm{F}_{6}^{\left(1;2,0\right)} & \equiv & \frac{1}{\sigma^{4}}\varepsilon_{abcd}\,{}^{4}\!R^{ae}\nabla^{b}\phi\nabla^{f}\phi\nabla^{c}\nabla_{e}\phi\nabla^{d}\nabla_{f}\phi,\\
	\bm{F}_{7}^{\left(1;2,0\right)} & \equiv & \frac{1}{\sigma^{6}}\varepsilon_{abcd}\,{}^{4}\!R_{ef}^{\phantom{ef}cd}\nabla^{m}\phi\nabla^{n}\phi\nabla^{e}\phi\nabla^{a}\phi\nabla^{f}\nabla_{m}\phi\nabla^{b}\nabla_{n}\phi,\\
	\bm{F}_{8}^{\left(1;2,0\right)} & \equiv & \frac{1}{\sigma^{6}}\varepsilon_{abcd}\,{}^{4}\!R_{ef}^{\phantom{ef}cm}\nabla^{a}\phi\nabla^{e}\phi\nabla_{m}\phi\nabla^{n}\phi\nabla^{b}\nabla_{n}\phi\nabla^{d}\nabla^{f}\phi,
	\end{eqnarray}
	\begin{eqnarray}
	\bm{F}_{1}^{\left(2;0,0\right)} & \equiv & \varepsilon_{abcd}\,{}^{4}\!R_{ef}^{\phantom{ef}cd}\,{}^{4}\!R^{abef},\\
	\bm{F}_{2}^{\left(2;0,0\right)} & \equiv & \frac{1}{\sigma^{2}}\,\varepsilon_{abcd}\,{}^{4}\!R_{ef}^{\phantom{ef}cd}\,{}^{4}\!R_{\phantom{abf}m}^{abf}\nabla^{e}\phi\nabla^{m}\phi,\\
	\bm{F}_{3}^{\left(2;0,0\right)} & \equiv & \frac{1}{\sigma^{2}}\,\varepsilon_{abcd}\,{}^{4}\!R_{ef}^{\phantom{ef}cd}\,{}^{4}\!R_{\phantom{efb}m}^{efa}\nabla^{b}\phi\nabla^{m}\phi,\\
	\bm{F}_{4}^{\left(2;0,0\right)} & \equiv & \frac{1}{\sigma^{2}}\,\varepsilon_{abcd}\,{}^{4}\!R_{ef}^{\phantom{ef}cd}\,{}^{4}\!R^{ae}\nabla^{b}\phi\nabla^{f}\phi,\\
	\bm{F}_{5}^{\left(2;0,0\right)} & \equiv & \frac{1}{\sigma^{4}}\,\varepsilon_{abcd}\,{}^{4}\!R_{ef}^{\phantom{ef}cd}\,{}^{4}\!R^{amen}\nabla^{b}\phi\nabla^{f}\phi\nabla_{m}\phi\nabla_{n}\phi.
	\end{eqnarray}
and
	\begin{equation}
	\bm{F}_{4}^{\left(1;0,1\right)}\equiv\frac{1}{\sigma^{4}}\varepsilon_{abcd}\,{}^{4}\!R_{ef}^{\phantom{ef}cd}\nabla^{a}\phi\nabla^{e}\phi\nabla^{m}\phi\nabla^{b}\nabla^{f}\nabla_{m}\phi,\label{F101_4}
	\end{equation}

%

\end{document}